\documentclass[12pt]{iopart}

\usepackage{iopams}  

\expandafter\let\csname equation*\endcsname\relax
\expandafter\let\csname endequation*\endcsname\relax

\usepackage{amssymb,amsmath,graphicx,color}
\usepackage{physics}
\usepackage[T1]{fontenc}
\usepackage{bm}
\usepackage{bbm}
\usepackage[tight]{subfigure}
\usepackage[export]{adjustbox}
\usepackage{enumerate}
\usepackage{braket}
\usepackage{appendix}
\usepackage{fancyhdr}
\usepackage[
  bookmarks=true,
  colorlinks,
  linkcolor=blue,
  citecolor=blue,
  urlcolor=blue,
  plainpages=false,
  pdfpagelabels,
  final,
  breaklinks=true
]{hyperref}

\usepackage[square,numbers,sort&compress]{natbib}

\usepackage{multirow}
\usepackage{rotating,booktabs}
\usepackage[verbose]{placeins}
\usepackage[dvipsnames]{xcolor}

\begin{document}

\title[]{Strong laser physics, non--classical light states and quantum information science}

\author{U. Bhattacharya$^{1}$, Th. Lamprou$^{2,3}$, A. S. Maxwell$^{4}$, A. Ordóñez$^{1}$, E.~Pisanty$^{5}$, J. Rivera-Dean$^{1}$, P. Stammer$^{1}$, M. F. Ciappina$^{6,7,8}$, M. Lewenstein$^{1,9}$ and P. Tzallas$^{2,10*}$ }

\address{$^{1}$ ICFO-Institut de Ciencies Fotoniques, The Barcelona Institute of
Science and Technology, Castelldefels (Barcelona) 08860, Spain}
\address{$^{2}$ Foundation for Research and Technology-Hellas, Institute of Electronic Structure \& Laser, GR-70013 Heraklion (Crete), Greece}
\address{$^{3}$ Department of Physics, University of Crete, P.O. Box 2208, GR-71003 Heraklion (Crete), Greece}
\address{$^{4}$ Department of Physics and Astronomy, Aarhus University, DK-8000 Aarhus C, Denmark}
\address{$^{5}$ Department of Physics, King's College London, WC2R 2LS London, United Kingdom}
\address{$^{6}$ Department of Physics, Guangdong Technion - Israel Institute of Technology, 241 Daxue Road, Shantou, Guangdong, China, 515063}
\address{$^{7}$ Technion - Israel Institute of Technology, Haifa, 32000, Israel}
\address{$^{8}$ Guangdong Provincial Key Laboratory of Materials and Technologies for Energy Conversion, Guangdong Technion - Israel Institute of Technology, 241 Daxue Road, Shantou, Guangdong, China, 515063}
\address{$^{9}$ ICREA, Pg. Lluís Companys 23, 08010 Barcelona, Spain}
\address{$^{10}$ ELI-ALPS, ELI-Hu Non-Profit Ltd., Dugonics tér 13, H-6720 Szeged, Hungary}

\address{$^{*}$E-mail: ptzallas@iesl.forth.gr}

\vspace{10pt}
\begin{indented}
\item[]\today
\end{indented}

\begin{abstract}
 Strong laser physics is a research direction that relies on the use of high-power lasers and has led to fascinating achievements ranging from relativistic particle acceleration to attosecond science. On the other hand, quantum optics has been built on the use of low photon number sources and has opened the way for groundbreaking discoveries in quantum technology, advancing investigations ranging from fundamental tests of quantum theory to quantum information processing. Despite the tremendous progress, until recently these directions have remained disconnected. This is because, the majority of the interactions in the strong-field limit have been successfully described by semi-classical approximations treating the electromagnetic field classically, as there was no need to include the quantum properties of the field to explain the observations. The link between strong laser physics, quantum optics, and quantum information science has been developed in the recent past. Studies based on fully quantized and conditioning approaches have shown that intense laser--matter interactions can be used for the generation of controllable entangled and non-classical light states. This achievement opens the way for a vast number of investigations stemming from the symbiosis of strong laser physics, quantum optics, and quantum information science. Here, after an introduction to the fundamentals of these research directions, we report on the recent progress in the fully quantized description of intense laser--matter interaction and the methods that have been developed for the generation of non-classical light states and entangled states. Also, we discuss the future directions of non-classical light engineering using strong laser fields, and the potential applications in ultrafast and quantum information science.

\end{abstract}

\maketitle
\tableofcontents

\pagestyle{fancy}
\fancyhf{} 
\setlength{\headheight}{16.0pt} 
\addtolength{\topmargin}{-2.0pt}
\renewcommand{\headrulewidth}{0pt} 
\fancyhead[L]{\it \nouppercase{\leftmark}}
\fancyhead[R]{\thepage}

%
\vspace{2pc}

\section {Introduction}

Until the end of the '50s, natural (incoherent) light was the main available source, with a well-formulated quantum description, that had been successfully used in the majority of investigations exploiting light--matter interactions. Because of this, the quantum description of a classically oscillating electromagnetic field (as given by classical electrodynamics) remained unexploited. This changed at the end of '50s and the beginning of '60s with three groundbreaking discoveries: the Hanbury Brown and Twiss experiment (conducted using incoherent light) \cite{HBT56}, the laser invention by T. H. Maiman (and earlier maser) (cf. \cite{Siegman-book} and references therein) and the formulation of the quantum theory of optical coherence and the invention of coherent states of light (used for the description of a classically oscillating field) by Roy Glauber and George Sudarshan (cf. \cite{glauber1963coherent, glauber1963quantum, sudarshan1963equivalence}). These achievements were poised to drastically speed up scientific progress and provide countless revolutionary discoveries moving from Nobel-prize winning experiments, into cross-disciplinary fields of basic research and technology \cite{Chu1998, CohenTannoudji1998, Phillips1998, CornelWieman2002, Ketterle2002, Glauber2006, Hall2006, Hansch2006, Haroche2013, Wineland2013, Ashkin2018, Mourou2019, Strickland2019, Aspect2022, Zeilinger2022, Clauser2022}. This scientific progress is largely based on the development of the research directions of quantum optics (QO), and strong laser physics (SLP), the latter by making possible the progress in laser technology and especially in pulsed laser sources \cite{Mourou2019, Strickland2019}.

In particular, QO is based on the quantum description of the electromagnetic radiation and largely on the formulation of the quantum theory of optical coherence \cite{glauber1963coherent, glauber1963quantum, sudarshan1963equivalence, Mandel&Wolf, Vogel&Welsch, Walls&Milburn}. A key aspect is the measurement of light intensity fluctuations and the characterization of the quantum state of the electromagnetic field, typically achieved by photon statistics measurements, the measurement of photon correlation functions \cite{glauber1963coherent, glauber1963quantum, HBT56, Kimble1977}, and by means of quantum tomography methods \cite{BSM97, LR09, Bachor_book_2019, Ariano2002, Blume2010, Auria2009, Vogel1989}. In the case of light--matter interaction, the majority of the studies are performed using weak electromagnetic fields (low photon number light sources) where the interaction is described by fully quantized theories, with the research, for many years, being focused on single or few body phenomena, trying to understand non-perturbative aspects of atom-light interaction (for a recent overview see \cite{Booklarson}). 

QO is at the core of quantum technologies \cite{Acin2018, Walmsay2015, Deutsch2020}. This is because light is considered as an ideal resource for engineering quantum states (such as squeezed, Fock, optical Schr\"{o}dinger ``cats'') \cite{BSM97, Leonhardt_book_1997, Andersen2016, Walls1983, Loudon1987, Wakui2007, Kimble1977, Diedrich1987, Michler2000,  Gadway2009, Magnitskiy2015, Takesue2004, Valles2014, Aspect2022, Zeilinger2022, Clauser2022, Haroche2013, Wineland2013, Zavatta2004, Ourjoumtsev2006, Hacker2019, Sychev2017, LCP21, RLP22} with non-classical properties.  These light states have some very distinct features and advantages compared to classical light. They can provide reduced noise and the notable feature of quantum correlations. 
Furthermore, light does not suffer as much from decoherence, in comparison to matter particles, when scaling to macroscopic sizes.
This is because the electromagnetic environment at optical frequencies can be considered as vacuum (i.e., as an environment with no particles) and thus relatively decoherence-free. These notable properties made non-classical light a key element for the emergence of new quantum technologies. Nowadays, the generation of non-classical and entangled light states has a vital role in quantum technologies as they offer a unique resource in a vast variety of investigations in quantum information science ranging from fundamental tests of quantum theory, quantum sensing, teleportation, communication, cryptography, visual science to high-precision interferometry applied for the detection of gravitational waves \cite{ Aspect2022, Zeilinger2022, Clauser2022, Haroche2013, Wineland2013, Acin2018, Walmsay2015, Deutsch2020, Gilchrist2004, Giovannetti2004, Giovannetti2011, Jouguet2013, Lloyd1999, Ralph2003, Joo2011, Schnabel2017}. However, despite the progress achieved so far, the applicability of the majority of the existing non-classical light sources is restricted by their low photon number, while the development of new schemes for the generation of high photon number non-classical and entangled light states is considered challenging.

On the other hand, SLP relies on the use of strong laser fields and largely on models employing semi-classical approximations, where the electromagnetic field is treated classically. The enhancement of laser power came almost a decade after the laser invention \cite{Maiman1960}, with the development of nanosecond (ns) and picosecond (ps) pulsed laser sources \cite{McClung1962, DeMaria1966}. By using these sources, the observation of non-linear processes was made possible. Due to the relatively low electric field of these laser pulses (compared to the Coulomb field of the atomic potential), these interactions were successfully described by the lowest-order perturbation theory using multi--photon absorption processes \cite{Delone2000}. At the end of 80s, the development of femtosecond (fs) \cite{Moulton1986, Fork1987} pulsed laser sources, combined with the pioneer development of the chirped-pulse amplification (CPA) technique \cite{Mourou2019, Strickland2019}, opened the way for time-resolved experiments in molecules \cite{Zewail1999} and investigations beyond the perturbative regime. In particular, high-power fs lasers allowed the scientific community to explore non-linear interactions in the strong field limit where the laser-electric field strength is comparable to, or even stronger than, the atomic potential \cite{protopapas_atomic_1997, kulander_dynamics_1993, corkum_plasma_1993, lewenstein-theory-1994}. Such interactions have opened the way for studies ranging from relativistic particle acceleration (\cite{Mourou2019} and references therein), to high harmonic generation \cite{ferray-multiple-1988, Li1989}, high-resolution spectroscopy in the extreme ultraviolet (XUV) \cite{Gohle2005, Cingoz2012} and attosecond physics (AP) \cite{Krausz2009, Corkum2007, Reduzzi2015, Borrego2022, Chatziathanasiou2017, Orfanos2019}. The interaction of atoms with strong laser pulses is at the core of the investigations in SLP and ultrafast science. Central to these studies, is the process of high harmonic generation where the low frequency photons of a driving laser pulse, after the interaction with atoms, are up-converted into photons of higher frequencies in the extreme ultraviolet spectral range. These interactions have been successfully described by classical or semi-classical strong field approximations, namely the three-step model \cite{kulander_dynamics_1993, corkum_plasma_1993, lewenstein-theory-1994, ABC19}, treating the electromagnetic field classically and ignoring its quantum nature. Because of this, advantages emerging from the connection of QO with SLP remained unexploited. By connecting QO with SLP, we can start by answering the following fundamental questions: (a) \emph{What is the back-action of the interaction on the driving field?} and (b) \emph{What is the quantum state of the radiation after the interaction with matter?} The importance of providing an answer to these two questions is directly associated with the applicability of SLP to quantum technologies. Several groups have provided early attempts to approach these problems \cite{LaGattuta2012, Wang2012, Yangaliev2020, Gauthey1995, Gombkoto2020, Varro2020, Gao2000, Hu2008, Fu2001, Kuchiev1999, Usachenko2002, Bogatskaya2017, Burenkov2010, Gorlach2020}. However, despite the important information provided by these studies, none of these efforts have addressed the above questions. This has recently changed. Theoretical and experimental investigations \cite{LCP21, RLP22, SRM2023}, conducted using fully quantized approaches in the strong-field limit, have linked SLP and ultrafast science with QO. This has been achieved by showing that the intense laser-matter interactions and conditioning approaches \cite{GTK2016, TKG17, TKD19, Sta22} on the process of high-harmonic generation, can be used for the generation of high photon number non-classical and entangled light states with controllable quantum features \cite{RLP22, SRM2023, SRL22, RSP21, RSM22}. These notable properties can be used for the development of a new class of non--classical and entangled states, advancing investigations in quantum technologies.

Here, after an introduction to the fundamentals of quantum optics and intense laser--matter interactions, we will discuss a fully quantized description of the latter. We will put emphasis on quantum electrodynamics (QED) of intense laser--atom interactions and the methods developed for engineering high photon number non-classical and entangled light states with controllable quantum features.  Then, we will provide potential directions towards the development of a new class of non--classical light sources using strongly laser-driven complex materials, and finally we will emphasize the generation of high photon number non--classical and massively entangled light states and the potential applications in ultrafast and quantum information (QI) science.

\section{Fundamentals of quantum optics and non-classical light engineering}\label{Section:2}
QO is a major research direction in the area of atomic, molecular and optical physics, which has been discussed in an enormous amount of books and articles published over the years (c.f.\ \cite{Mandel&Wolf, Vogel&Welsch, Walls&Milburn, Gerry_book_2005, Leonhardt_book_2010, Scully_book_1997, Lambropoulos_book_2007}). However, for reasons of completeness and for introducing the terminology and definitions relevant to the present article, here, we will discuss a few of the fundamentals of this research direction. 

\subsection{Quantum harmonic oscillator}\label{Section:fund:QO}
The foundations of QO are based on the quantization of the electromagnetic field which, when expanded into its frequency modes, can be written as a set of independent quantum harmonic oscillators (see c.f.~\cite{Scully_book_1997,Gerry_book_2005,Vogel&Welsch} for a detailed derivation), each of them having a frequency $\omega$. Thus, for a single mode, we can write the corresponding Hamiltonian as $H = \tfrac{1}{2}(\hat{\bar{p}}^2 + \omega^2 \hat{\bar{x}}^2) = \hbar \omega (\hat{a}^\dagger \hat{a} + 1/2)$, where $\hat{\bar{x}}$ and $\hat{\bar{p}}$ are the canonical operators of the harmonic oscillator, analog to the position and momentum operators of a particle in a harmonic potential, which satisfy the commutation relation $[\hat{\bar{x}},\hat{\bar{p}}] = i\hbar$; and $\hat{N} = \hat{a}^\dagger\hat{a}$ is the photon number operator, with $\hat{a}^\dagger$ and $\hat{a}$ the creation and annihilation operators, respectively, which satisfy $[\hat{a},\hat{a}^\dagger] = 1$. The canonical variables $\hat{\bar{x}}$ and $\hat{\bar{p}}$ are related to the creation and annihilation operators by $\hat{\bar{x}} = \sqrt{\hbar/(2\omega)}(\hat{a} + \hat{a}^\dagger)$ and $\hat{\bar{p}} = i\sqrt{\hbar\omega/2}(\hat{a}^\dagger - \hat{a})$. However, instead of working with the canonical variables $\hat{\bar{x}}$ and $\hat{\bar{p}}$, in quantum optics we usually work with dimensionless operators referred to as \emph{quadrature operators}, $\hat{x}$ and $\hat{p}$, which are related to the creation and annihilation operators as $\hat{x} = (\hat{a}+\hat{a}^\dagger)/\sqrt{2}$ and $\hat{p} = (\hat{a}-\hat{a}^\dagger)/(i\sqrt{2})$.

The eigenstates of the Hamiltonian presented before, are the so-called Fock states $\ket{n}$, with $n \in \mathbbm{N}$ and eigenvalue $\hbar\omega(n+1/2)$. Thus, since they are eigenstates of the photon number operator $\hat{N}$, Fock states (also known as photon number states) have a well-defined photon number, where the state that has $n=0$ photons, i.e. $\ket{0}$, defines the so-called \emph{vacuum} state. In the quadrature $x$-representation, a Fock state can be written as $\bra{x}\ket{n}=\psi_{n}(x)=\psi_{0}(x) H_{n}(x)/\sqrt{2^n n!}$, where $H_{n}(x)$ is the $n$th-order Hermite polynomial and $\psi_{0}(x)=\exp(-x^2/2)/\pi^{1/4}$ is the wavefunction of the vacuum state.

Finally, the electric field operator corresponding to a single mode of the electromagnetic field is given by $\hat{E}(t) = \sqrt{\hbar \omega/(\epsilon_0 V)} (\hat{a} e^{-i\omega t} + \hat{a}^\dagger e^{i\omega t})$, with $V$ the quantization volume and $\epsilon_0$ the vacuum permittivity. In terms of the quadrature operators, it reads as $\hat{E}(t) = \sqrt{2\hbar \omega/(\epsilon_0 V)} (\cos(\omega t) \hat{x} + \sin(\omega t)\hat{p})$. We will use these expressions further on in the text.

\subsection{Coherent states of light}\label{Sec:Coherent:States}

The quantum description of a classically oscillating field becomes feasible with the formulation of coherent states of light. A single-mode classical electromagnetic field is often described as a wave with a well-defined amplitude and phase. However, in the quantum theory of radiation, the amplitude and the phase are conjugate variables, and therefore cannot be determined with arbitrary accuracy in the same experiment. Coherent (or Glauber) states of light, are states for which the product of the variances of these two quantities reaches its lower limit, i.e. the fluctuations in the two quadratures are equal, and minimize the uncertainty product given by Heisenberg's uncertainty relation $\Delta x \Delta p = \frac {1}{2}$ with $\Delta x = \Delta p = \frac {1}{\sqrt{2}}$. Therefore, coherent states provide the optimal description of a classical field. They are typically denoted as $\ket{\alpha}$, with Greek letters, and can be expanded in the Fock state basis as
\begin{align}
\ket{\alpha} = e^{-\abs{\alpha}^2 /2} \sum_{n=0}^\infty \frac{\alpha^n}{\sqrt{n!}} \ket{n}, 
\end{align}
with  $\alpha=\lvert\alpha\rvert\exp(i\theta)$, the complex amplitude, such that the mean-photon number is given by $\expval{n} = \bra{\alpha}\hat{N}\ket{\alpha}=|\alpha|^2$. These states are eigenstates of the annihilation operator, i.e. $\hat{a} \ket{\alpha}=\alpha \ket{\alpha}$, and can be generated by coupling a classically oscillating current, $\vb{J}(\vb{r},t)$, to the vector potential operator of an electromagnetic field mode, $\vb{A}(\vb{r},t)$, so that the Hamiltonian describing the interaction is $\hat{H} = \int \dd^3 r \vb{J}(\vb{r},t) \cdot \hat{\vb{A}}(\vb{r},t)$. The unitary evolution associated to this Hamiltonian can be written in terms of the displacement operator $\hat{D}(\alpha) = \exp(\alpha\hat{a}^\dagger - \alpha^*\hat{a})$, which induces a shift in phase-space of the corresponding initial state, with the amplitude $\alpha$ proportional to the Fourier component of the classical current with respect to the mode with frequency $\omega$. If, for instance, we consider the initial state to be the vacuum state $\ket{0}$, then we get a coherent state of amplitude $\alpha$, i.e., $\ket{\alpha} = \hat{D}(\alpha)\ket{0}$. Equivalently, in terms of the quadrature representation, we can write $\alpha=(x_{0}+ip_{0})/\sqrt{2}$, with $x_0 =\mel{\alpha}{\hat{x}}{\alpha}$ and $p_0 =\mel{\alpha}{\hat{p}}{\alpha}$, the displacement operator as $\hat{D}=\exp(ip_{0}\hat{x}-ix_{0}\hat{p})$ and its wavefunction in the $x$-representation as $\bra{x}\ket{\alpha}=\psi_{\alpha}(x)= \psi_{0}(x-x_{0}) \exp(ip_{0}x-ip_{0}x_{0}/2)$, where $\psi_0(x-x_{0})=\pi^{-1/4}e^{-(x-x_{0})^2/2}$ is the wavefunction of the shifted vacuum state. The time-evolution of coherent states under free-field dynamics, i.e. dictated by the Hamiltonian $\hat{H} = \hbar\omega(\hat{a}^\dagger \hat{a} + 1/2)$, is given by $\ket{\alpha(t)}=\exp(-i\hat{H}t/\hbar)\ket{\alpha}=\ket{\alpha \exp(-i\omega t)}$, from which it stems that coherent states remain coherent under free-field evolution. Thus, in the quadrature $x$-representation $\bra{x}\ket{\alpha(t)}=\psi_{\alpha}(x,t)$, the obtained wavefunction describes a Gaussian wavepacket, unchanged in time, which oscillates with the frequency of the field following a motion like a classical particle in a harmonic oscillator potential. 

If, by using the above, we compute the expectation value of the electric field operator, we get $\bra{\alpha}\hat{E} (t) \ket{\alpha}=|\alpha| \cos(\omega t -\theta)$, which is the expression of a monochromatic electromagnetic field obtained in classical electromagnetism. For this reason coherent states are referred to as ``classical'' or ``quasi-classical''. Fields described by a statistical mixture of coherent states, such as thermal fields, are considered as classical as well, while any state that cannot be described by this kind of mixtures is typically regarded as ``non-classical''.

\subsection{Quantum state characterization} 
The characterization of the quantum state of light is a large chapter in QO and it is practically impossible to address it in its entirety in a single section of a manuscript. Some of the most useful methods that are commonly used to characterize and distinguish the classical from the non--classical light fields are based on the measurement of the (I) $q$th order correlation functions $g^{(q)} (\tau)$ \cite{glauber1963coherent, glauber1963quantum, Glauber2006}, (II) photon number distribution, and (III) Wigner function (or Q-, P-functions) in phase space \cite{Schleich_book_2001}. Each of these methods alone is sufficient, but not necessary, to distinguish classical from non--classical light (for details see Ref. \cite{Strekalov_book_2019} and references therein).

In method (I), the $g^{(q)}$ functions are used (with $q$=2 being the most commonly used) to characterize the statistics and the degree of coherence of an electromagnetic field. The $g^{(q)} (\tau)$ functions are typically obtained by interferometric approaches and $q$th order autocorrelation measurements, with $q$ being the order of nonlinearity and $\tau$ the time delay between the photon signals. In method (II), the photon number distribution, given by $P_{n}=\lvert\braket{n}{\psi}\rvert^2$ (where $\ket{\psi}$ is the quantum state of the field), can be obtained directly by photon statistics measurements. Regarding these methods, a coherent state depicts a Poissonian photon number distribution with $P_{n}=\lvert\braket{n}{\alpha}\rvert^2=e^{-|\alpha|^2}|\alpha|^{2n}/n!=e^{-\langle n\rangle}\langle n\rangle^{n}/n!$, and a normalized $g^{(2)}$ = 1. This refers to the case where photons randomly reach a detector. A light source is called super-Poissonian if the photon number distribution leads to $g^{(2)}(0) > 1$, and photon bunching is observed. This refers to the case where the photons have the tendency to reach the detector in bunches, i.e., more close in space (time) than the photons of the coherent state. It is noted that incoherent light also corresponds to this case with the photon number fluctuations to be determined by the coherence time of the light source. A non--classical light source with sub-Poissonian photon number distribution and $0< g^{(2)}(0) < 1$, shows photon antibunching signatures. This is a pure quantum effect which refers to the case where the photons have the tendency to reach the detector more equally and further away in space (time) than those of a coherent state. 

\begin{figure*}
    \centering
    \includegraphics[width=0.3 \linewidth]{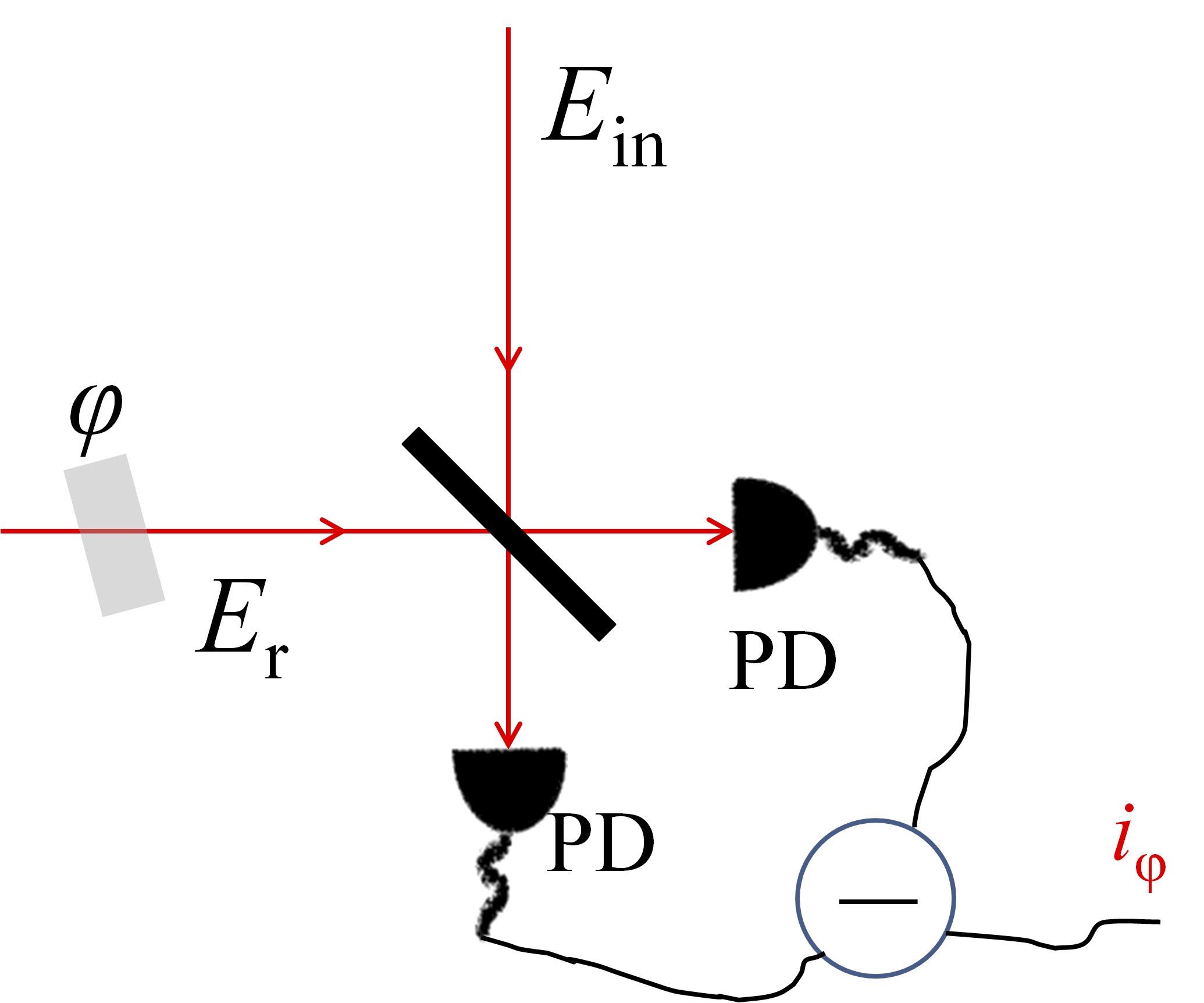}
    \caption{Homodyne detection and quantum tomography approach. E$_{in}$ is the state of the field to be characterized, while $E_{\text{r}}$ is the field of a local oscillator $E_{\text{r}}$ (with $E_{\text{r}}\gg E_{\text{in}}$). BS is a beam splitter, and $\varphi$ is the controllable phase shift between the $E_{\text{in}}$ and $E_{\text{r}}$ field. Two identical photodiodes, labeled PD, use a balanced differential photodetection system, which provides, at each value of $\varphi$, the photocurrent difference $i_{\varphi}$. The characterization of the quantum state of light can be achieved by recording the value of $i_{\varphi}$ as a function of $\varphi$.}
    \label{fig:QT}
\end{figure*}

\begin{figure*}
    \centering
    \includegraphics[width=0.7 \linewidth]{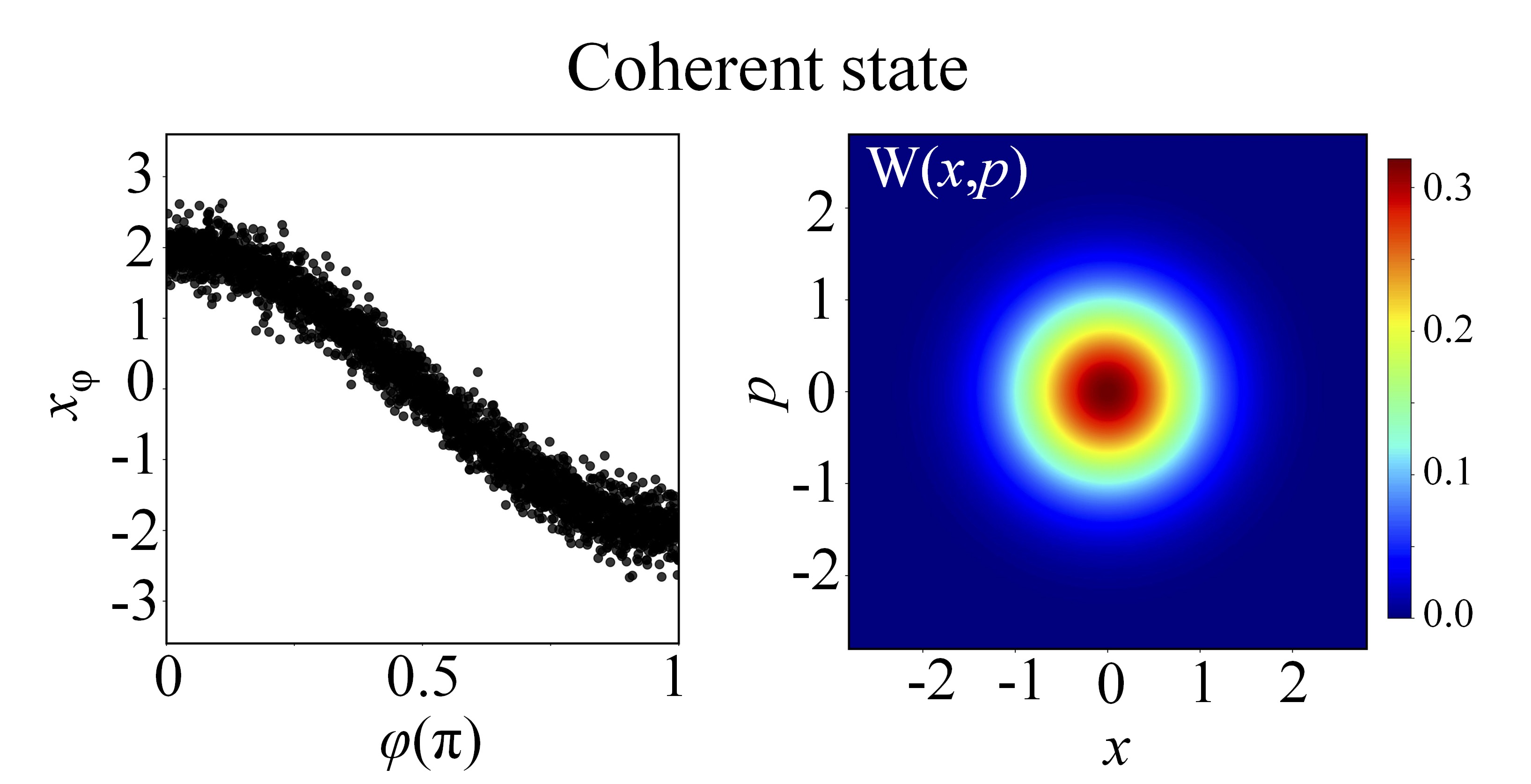}
    \caption{Homodyne trace and Wigner function of a coherent light state $\ket{\alpha}$. The left panel shows the calculated homodyne detection signal $x_{\phi}$, and the right panel the corresponding Wigner function $W(x,p)$ which has been centered at $|\alpha|=2$. The Figure has been reproduced from ref. \cite{SRM2023}.}
    \label{fig:W_coherent}
\end{figure*}

In method (III), the Wigner function can be obtained by implementing a homodyne detection technique and quantum tomography method \cite{BSM97, LR09, Bachor_book_2019}. This method is considered as one of the most complete ways of characterizing a light state. A simplified optical layout of the technique implemented by means of an interferometric approach, is shown in Fig.~\ref{fig:QT}. 
$E_{\text{in}}$ is the field of the state to be characterized. The $E_{\text{in}}$ field is spatiotemporally overlapped on a beam splitter (BS) with the field of a local oscillator $E_{\text{r}}$ (with $E_{\text{r}}\gg E_{\text{in}}$). The $E_{\text{r}}$ field comes from the 2nd branch of the interferometer which introduces a controllable phase shift $\varphi$ between the $E_{\text{in}}$ and $E_{\text{r}}$ fields. The fields after the BS are detected by a balanced differential photodetection system consisting of two identical photodiodes (PD). This provides at each value of $\varphi$ the photocurrent difference $i_{\varphi}$. The characterization of the quantum state of light can be achieved by recording the value of $i_{\varphi}$ as a function of $\varphi$. These values are directly proportional to the measurement of the electric field operator $\hat{E}_{\text{in}} (\varphi)\propto \hat{x}_{\varphi} \propto \cos(\varphi) \hat{x}+ \sin(\varphi) \hat{p}$ and are used for the reconstruction of the Wigner function via Radon transformation \cite{Herman80}. Experimentally, this is all that one has to do in order to characterize the light state. This is because repeated measurements of $\hat{x}_{\varphi}$ at each $\varphi$ provides the probability distribution $P_{\varphi}(x_{\varphi})=\langle{x_{\varphi}}|{\hat{\rho}}|{x_{\varphi}}\rangle$ of its eigenvalues $x_{\varphi}$, where $\hat\rho\equiv\dyad{\phi}$ is the density operator of the light state to be characterized and $|{x_{\varphi}}\rangle$ the eigenstate with eigenvalue $x_{\varphi}$. The density matrix $\hat\rho$, which provides complete information about the light state, can be obtained in the Fock basis by calculating the matrix elements $\rho_{nm}$ using an iterative {\it Maximum--Likelihood} procedure beautifully described in Ref.~\cite{Lvovsky_MaxLik_alg2004}. Having determined these values, the mean photon number of the light state can be obtained by the diagonal elements $\rho_{nn}$ of the density matrix $\hat \rho$, and the relation $\langle{n}\rangle=\sum n\rho_{nn}$. Fig.~\ref{fig:W_coherent} shows a calculated homodyne trace (left panel) and the reconstructed Wigner function (right panel) of a coherent state.
The Wigner function, which is the Wigner function of the vacuum state shifted by $(x_{0}, p_{0})$, depicts a Gaussian distribution of the form $W(x,p) = \frac{1}{\pi} \exp[-(x-x_0)^2-(p-p_0)^2]$ which in a more compact form reads,
\begin{equation}\label{Eq:Wigner:Coher:Compact}
\begin{aligned}
W(\beta) = \frac{2}{\pi} e^{-2|\beta - \alpha|^{2}}.
 \end{aligned}
\end{equation} 
In the above equation we have used the transformation $\it x \equiv \Re[\beta-\alpha]$ and $\it p \equiv \Im[\beta-\alpha]$ (where $\beta=(x+ip)/\sqrt{2}$ is a variable), which centers the Wigner function at the origin for $\beta=\alpha$, $x=0$ and $p=0$.

\subsection{Non-classical light states}

Any state that cannot be described by a mixture of coherent states, or provides a Wigner function that depicts negative values and/or non-Gaussian distributions, is considered non-classical. Such states include Fock states, squeezed states, and coherent state superpositions, e.g., optical Schrödinger ``cat'' states. Because these states are central to quantum technologies, in the following subsections we provide some of their main features.

\subsubsection{Photon number (Fock) states}
\begin{figure*}
    \centering
    \includegraphics[width=0.7 \linewidth]{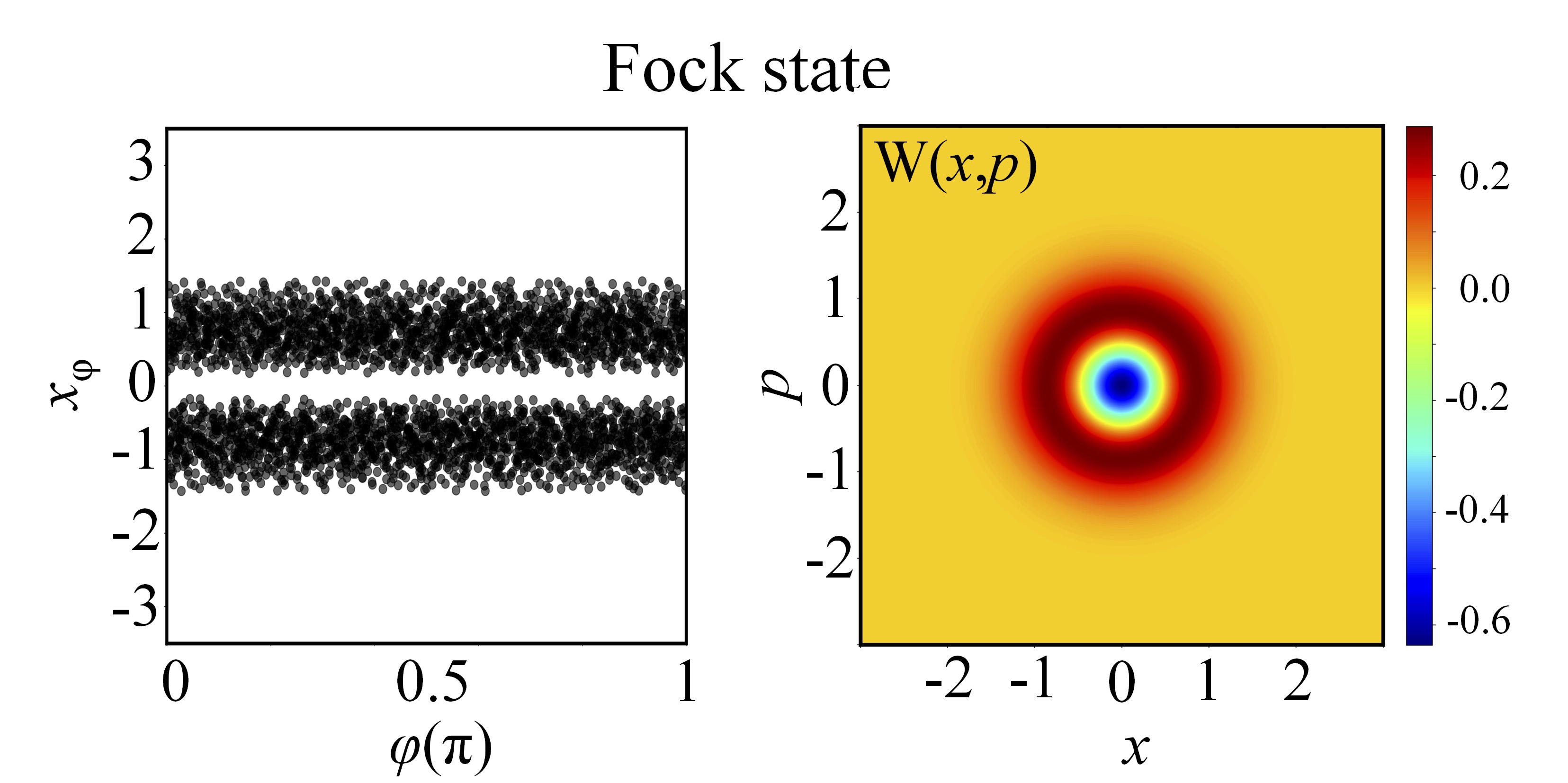}
    \caption{Homodyne trace and Wigner function of a Fock state $\ket{n}$ with $n=1$. The left panel shows the calculated homodyne detection signal $x_{\phi}$, and the right panel the corresponding Wigner function $W(x,p)$.}
    \label{fig:W_Fock}
\end{figure*}

\hfill \break
From an experimental perspective, Fock states are delivered by single--photon sources, which have been developed by means of single atom and molecule transitions \cite{Kimble1977, Moerner1989}, ion--traps \cite{Diedrich1987}, defect centers in solid state materials \cite{Aharonovich2016, Castelletto2014}, quantum dots \cite{Michler2000}, 2D materials \cite{Branny2017}, etc. Nowadays, the most commonly used source is based on a non-linear parametric down-conversion process \cite{Gadway2009, Magnitskiy2015, Takesue2004, Valles2014}, where a high energy photon, after interacting with a non-linear medium, is converted into a pair of lower-energy photons. Fock states depict photon anti--bunching properties with a sub-Poissonian photon number distribution, confined at the photon number $n$. The Wigner function has a ring--shape structure, centered at $(x=0,p=0)$, and depicts negative values. It has the form $W_{n}(x,p) = \frac{(-1)^n}{\pi} \exp(-x^2-p^2)L_{n}(2x^2+2p^2)$ which, according to the transformation used in Eq.~\eqref{Eq:Wigner:Coher:Compact}, reads, 
\begin{equation} \label{Eq:WignerFock}
\begin{aligned}
W_{n}(\beta) = \frac{2}{\pi} (-1)^n e^{-2|\beta|^{2}} L_{n}(4|\beta|^2),
 \end{aligned}
\end{equation}
where $L_{n}$ are the $n$--order Laguerre polynomials. Fig.~\ref{fig:W_Fock}, shows an example of a calculated homodyne trace (left panel) and the corresponding Wigner function (right panel) of the single photon Fock state $\ket{1}$.

\subsubsection{Squeezed light states}
\hfill \break
In contrast to coherent states of light, quantum fluctuations for squeezed states of light are not equally distributed between the field quadratures. They are minimum uncertainty states, which have reduced fluctuations (compared to the coherent or vacuum state) in one quadrature and increased in the other. They are produced by non--linear interactions, with the parametric down conversion process in a crystal being one of the most commonly used methods \cite{Andersen2016} (and references therein). This interaction can be described in terms of the Hamiltonian $\hat{H}=\hbar\chi(\hat{a}^2 - {{}\hat{a}^\dagger}^2)$, and its unitary evolution introduces the so-called \emph{squeezing operator}, given by $\hat{S}(k)=\exp[-\frac{k}{2}(\hat{a}^2 - {{}\hat{a}^\dagger}^2)]$. In this expression, $k=2i\chi \tau$ is the squeezing parameter, $\chi$ the non-linear coupling parameter and $\tau$ the interaction time. By applying $\hat{S}(k)$ to a vacuum state $\ket{0}$, we obtain the \emph{squeezed-vacuum} state $\ket{\text{SV}}=\hat{S}(k)\ket{0}$, while the displaced squeezed vacuum state $\ket{\text{DSV}}$ can be obtained by applying the displacement operator $\hat{D}(\alpha)$ on $\ket{\text{SV}}$, i.e. $\ket{\text{DSV}}=\hat{D}(\alpha)\hat{S}(k)\ket{0}$, whose wavefunction in $x$-representation is $\psi_{\text{DSV}}(x)= \pi^{-1/4}e^{-k /2} \exp[-e^{2k} \frac{(x-x_{0})^{2}}{2}+ipx-i\frac{x_{0}p_{0}}{2}]$. The Wigner function of these states corresponds to a Gaussian which has been `squeezed' along one of its quadratures and stretched in the other, which according to the transformation used in Eq.~\eqref{Eq:Wigner:Coher:Compact} reads, 
\begin{equation} \label{Eq:WignerSqueezed}
\begin{aligned}
W_{DSV}(\beta) = \frac{2}{\pi} e^{-2\abs{\frac{(\beta-\alpha)}{e^{-2k}}+\frac{(\beta^{*}-\alpha^{*})}{e^{2k}}}^{2}}.
 \end{aligned}
\end{equation}
\begin{figure*}
    \centering
    \includegraphics[width=0.7 \linewidth]{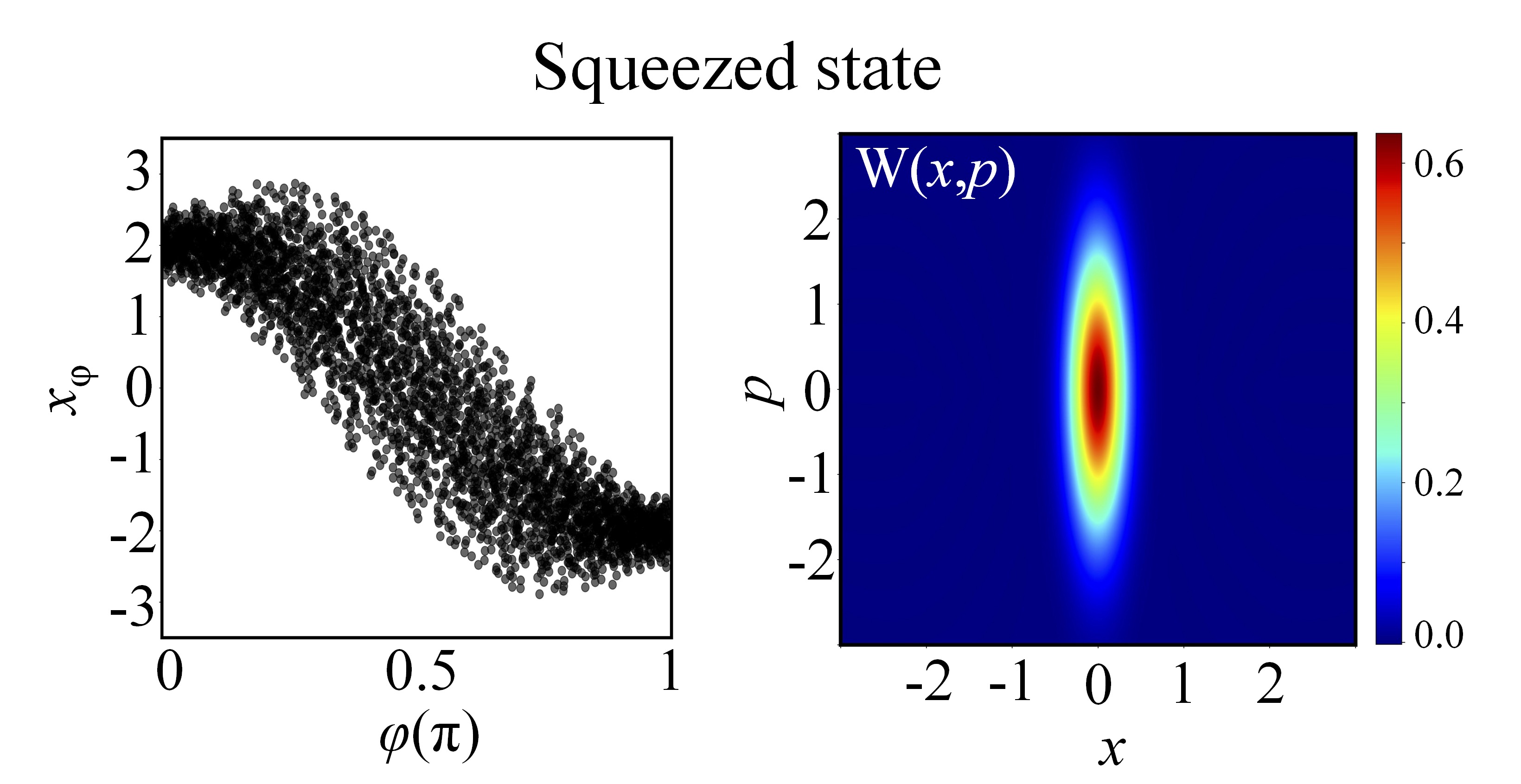}
    \caption{Homodyne trace and Wigner function of an amplitude squeezed state with $|a|=2$ and $k=0.8$. The left panel shows the calculated homodyne detection signal $x_{\phi}$, and the right panel the corresponding Wigner function $W(x,p)$ which has been centered at $|\alpha|=2$.}
    \label{fig:W_Squeezed}
\end{figure*}
Fig.~\ref{fig:W_Squeezed} shows an example of a calculated homodyne trace (left panel) and the corresponding Wigner function (right panel) of an amplitude squeezed state with $|\alpha|=2$ and $k=0.8$.

\subsubsection{Coherent state superpositions and optical ``cat'' states} 
\begin{figure*}
    \centering
    \includegraphics[width=0.7 \linewidth]{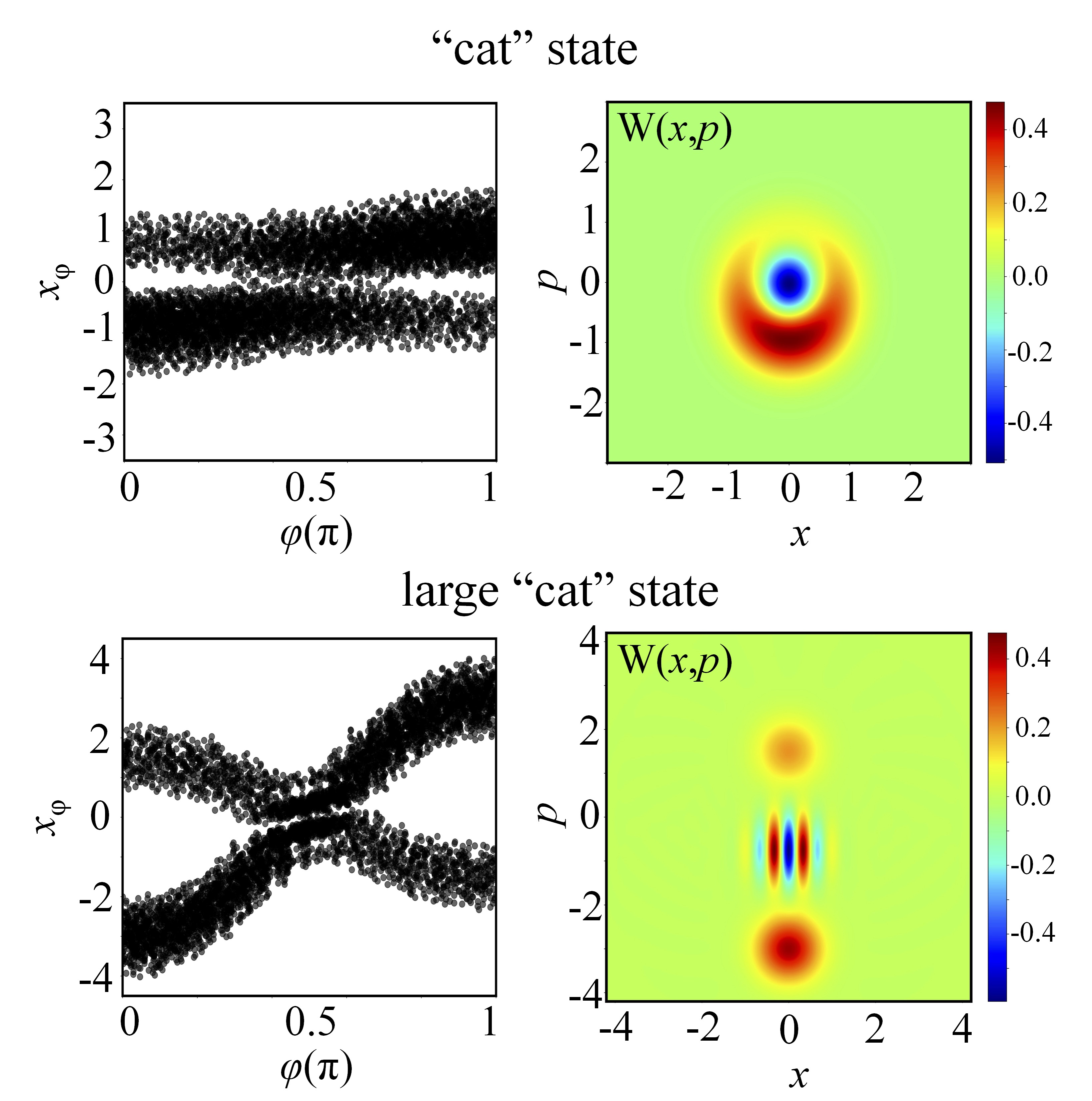}
    \caption{Examples of homodyne traces and Wigner functions of coherent state superpositions $\ket{\text{CSS}}=\xi_{1} \ket{\alpha_{1}}+ \xi_{2}\ket{\alpha_{2}}$. The left panels show the calculated homodyne detection signal $x_{\phi}$, and the right panels the corresponding Wigner function $W(x,p)$. (Upper panel):  A shifted optical ``cat'' state with $\xi_{1}=0.7$, $\xi_{2}=-1$, $\alpha_{1}=0.3$ and $\alpha_{2}=-0.6$. (Lower panel): A shifted large optical ``cat'' state with $\xi_{1}=0.7$, $\xi_{2}=-1$, $\alpha_{1}=1.5$ and $\alpha_{2}=-3$.}
    \label{fig:W_Cat}
\end{figure*}

\hfill \break
The superposition of coherent states that differ in amplitude and/or phase, can exhibit genuine quantum features without a classical counterpart \cite{Haroche_book_2006, Leonhardt_book_1997, Leonhardt_book_2010, Gerry_book_2005}. The generic form of a coherent state superposition (CSS) in a single field mode is given by $\ket{\text{CSS}} = \sum_i \xi_i \ket{\alpha_i}$, with $\alpha_i \neq \alpha_j \ \forall i\neq j$, and where $\xi_i$ is the corresponding probability amplitude. In the case where the superposition is composed by two coherent states of equal amplitude $\abs{\alpha}$ and opposite phase, the state takes the form $\ket{\text{cat}}_{\pm} = \frac{1}{N_{\pm}}(\ket{\alpha} \pm \ket{-\alpha})$, where $N_{\pm}=1/\sqrt{2(1\pm e^{-2|\alpha|^2})}$ are the corresponding normalization factors. The $\ket{\text{cat}}_{+}$ and $\ket{\text{cat}}_{-}$ contain only even and odd photon numbers, respectively, when expanded in the Fock basis, and are usually referred to as {\it even} and {\it odd} optical Schr\"{o}dinger ``cat'' states. This is because, within the optical domain, they resemble the states Schrödinger presented in his famous {\it Gedankenexperiment} with cats \cite{schrodinger1935gegenwartige}. This can be directly seen by the wavefunction in $x$-representation of this state, i.e.~$\psi_{\pm}(x) \propto e^{-(x-x_{0})^2/2}\pm e^{-(x+x_{0})^2/2}$, which shows two peaks, one at $+x_{0}$ (which would correspond to a \textit{dead cat}) and one at $-x_{0}$ (an \textit{alive cat}). Also, depending on the amplitude difference of the coherent states in the superposition, the states are named optical ``kitten'' and large optical ``cat'' states for small and large amplitude differences, respectively. In a more general case, when the superposition of two coherent states has the form,
\begin{equation}\label{eq:cat}
\ket{\text{CSS}} = \ket{\alpha_1} \pm \xi \ket{\alpha_2}, 
\end{equation}
we will refer to it as a \textit{shifted ``cat''} state, where $\xi=\bra{\alpha_{1}}\ket{\alpha_{2}}=e^{-\frac{1}{2}(|\alpha_1|^2+|\alpha_2|^2-2\alpha_{1}^{*}\alpha_2)}$ is the coupling of the two states. The Wigner function of an optical ``cat'' depicts a ring-shaped structure and with negative values in the region where the two coherent states overlap. For large optical ``cat'' states, the Wigner function depicts two pronounced Gaussian-like maxima at $(x_{i}, \pm p_{i})$, where $\pm x_i =\mel{\alpha_i}{\hat{x}}{\alpha_i}$ and $\pm p_i =\mel{\alpha_i}{\hat{p}}{\alpha_i}$, with a strong interference pattern with negative values in the region where the two coherent states overlap. In both cases, the negative values are associated with pure quantum interference effects. Applying the transformation used in Eq.~\eqref{Eq:Wigner:Coher:Compact}, the Wigner function of Eq.~\eqref{eq:cat} reads,
\begin{equation} \label{eq:WignerPlot}
\begin{aligned}
W(\beta)
			=\ &\dfrac{2}{\pi N}
			 \Big[ e^{-2\lvert\beta - \alpha - \chi\rvert^2}
			 + e^{-\lvert\chi\rvert^2}e^{-2\lvert\beta - \alpha\rvert^2}\\
			 &- \big(
			 		e^{2(\beta - \alpha)\chi^*}
			 		+ e^{2(\beta - \alpha)^*\chi}
			 	\big)
			 	e^{-\lvert\chi\rvert^2}e^{-2\lvert\beta - \alpha\rvert^2}
			 	\Big],
 \end{aligned}
\end{equation}
where $N = 1 - e^{-\lvert\chi\rvert^2}$ is the normalization factor for $\ket{\text{CSS}} = \ket{\alpha_1} - \xi \ket{\alpha_2}$ with $\ket{\alpha_1}=\ket{\alpha_2+\chi}$. Figures~\ref{fig:W_Cat}a and~\ref{fig:W_Cat}b, show two examples of the calculated homodyne trace (left panels), and the corresponding Wigner functions (right panels) of a shifted optical ``cat'' and a shifted large optical ``cat'' state, respectively. 

The optical ``cat'' states are considered as a resource for a vast variety of investigations in quantum technology and quantum information science. However, their generation is non-trivial and requires the implementation of sophisticated quantum optical protocols. The first observation of these states was achieved by the pioneering works of S.\ Haroche and D.~J.\ Wineland, where cat states were conduced by means of cavity QED and ion trap experiments \cite{Haroche2013, Wineland2013}. Later, and particularly from 2004 to 2017, fully optical methods in quantum state engineering were developed for the generation of optical ``cat'' and ``kitten'' states (cf. \cite{Zavatta2004, Ourjoumtsev2006, Ourjoumtsev2007, Dakna1997, Sychev2017}). These methods, were developed by means of linear optical elements and the use of Fock, coherent, and squeezed light states as primary sources. However, the low photon number ``cat'' states delivered by these sources restricts their applicability in many novel investigations in quantum technologies, which can be greatly advanced by the development of new schemes that can lead to the generation of high-photon-number optical ``cat'' states with controllable quantum features. Such schemes have recently been developed by means of intense laser--matter interactions. Specifically, recent theoretical and experimental investigations \cite{LCP21, RLP22, SRM2023}, conducted using fully quantized approaches in the strong-field limit, have shown that the intense laser--atom interactions and conditioning approaches \cite{GTK2016, TKG17, TKD19, Sta22} in the process of high-harmonic generation, can be used for the generation of high photon number non-classical and entangled light states with controllable quantum features \cite{RLP22, SRM2023, SRL22, RSP21, RSM22}. This matter will be further discussed in Sec. \ref{Sec:QED_Atoms}.

\subsubsection{Entangled states}
\hfill \break
So far we have only considered a single mode of the electromagnetic field, which can be described by a state $\ket{\psi_1}$, e.g. with a Fock, coherent or squeezed state. However, if we take into account an additional mode of the field, described by the state $\ket{\psi_2}$, we can now consider the total system of the field composite of the two modes. 
Since we now have two modes, which can, for instance, be different spatial or frequency modes, we can speak about a bipartite system in which each field mode represents one subsystem. The total state of the field, for example, is now given by the tensor product $\ket{\psi_1} \ket{\psi_2}$. However, the most generic case of the total state for the two field modes is expressed in a basis expansion for each mode 
\begin{align}
	\ket{\Psi} = \sum_{ij} c_{ij} \ket{i} \ket{j},
\end{align} 
where $c_{ij} = \bra{ij} \ket{\Psi}$ for the basis states $\ket{i}$ and $\ket{j}$ for the two modes, respectively.
The state $\ket{\Psi}$ is now said to be entangled, if it can not be written in a separable tensor product form, i.e. $\ket{\Psi} \neq \ket{\psi_1} \ket{\psi_2}$. The fact that the quantum formalism allows for entanglement \cite{horodecki2009quantum} initially caused interpretational difficulties by means of the famous work from Einstein, Podolsky, Rosen (EPR) and Schrödinger \cite{einstein1935can, schrodinger1935gegenwartige}, leading to questions on and study of the foundations of quantum theory \cite{horodecki2009quantum}.
Almost 90 years later, the concept of entanglement has evolved into a versatile tool and resource for modern quantum technologies \cite{reid2009colloquium}. 

For the purpose of this manuscript, we focus on the description of entangled states of different frequency modes of the field, each represented by a coherent state. Such entangled coherent states \cite{sanders2012review} will be found in the analysis of the process of HHG \cite{SRL22, Sta22, SRM2023} (see Sec. \ref{Sec:QED_Atoms}).
In addition to using the process of HHG, entangled coherent states can be generated using interferometer schemes \cite{sanders1992entangled} or when a coherent state superposition is impinging onto a beam splitter \cite{van2001entangled}.

\section{Fundamentals of intense laser-matter interactions: Semi-classical approaches}
\label{Sec:SLFP:SemiClas}
The tremendous progress in fs laser pulse engineering \cite{Nolte_book_2016} has led to the development of table-top laser systems, which nowadays can deliver high power laser pulses of duration down to few optical cycles, and carrier wavelength ranging from the visible and near-infrared (IR), up to mid-infrared (M-IR) spectral range \cite{Biegert2011}, and THz pulses \cite{Hebling2012}. 

\begin{figure*}
    \centering
    \includegraphics[width=0.85 \linewidth]{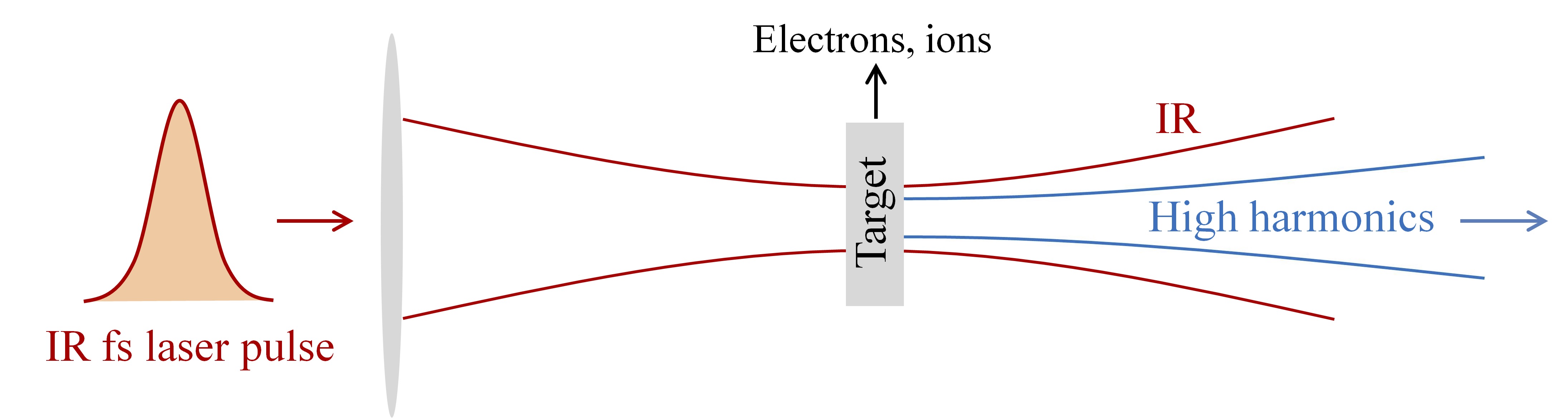}
    \caption{An oversimplified drawing of a typical approach used for investigating intense laser--matter interactions. An IR laser pulse is focused into a target area which contains the medium to be investigated (atoms, molecules, solids, etc.). Considering atoms as a target medium, the intensity of the driving laser pulse in the interaction area is typically $I_{L}>10^{14}$ W/cm$^2$ for noble gases. The medium length is smaller than the confocal parameter of the driving laser and thus experiences the same intensity along the propagation direction. The interaction products are electrons, ions and high harmonics emitted in the direction of the driving field.}
    \label{fig:Fig_Intense_laser_matter}
\end{figure*}

The high intensities achievable with these pulses have enabled the investigation of laser--matter interactions in the strong-field limit, where the laser-electric field strength is approaching the binding field inside the atoms themselves. Interactions in this intensity region led to the observation of many fascinating non-linear processes in all states of matter. Among these, is the intense laser--atom interaction and the process of high harmonic generation (HHG), which has opened the way for numerous investigations ranging from the high-resolution spectroscopy in XUV to AP. A typical arrangement used for investigating these interactions is shown in a simplified way in Fig.~\ref{fig:Fig_Intense_laser_matter}.
Briefly, a linearly polarized fs laser pulse is focused into a target. Considering atoms (noble gases) as a target medium, the intensity of the driving laser pulse in the interaction area is typically $I_{L}>10^{14}$ W/cm$^2$. The interaction leads to the generation of ions, photoelectrons, and high-harmonic photons emitted in the propagation direction of the laser field. Due to the non-linearity of the harmonic generation process the divergence of the harmonic beam is smaller than the driving field. Central in these studies is the measurement, and characterization of the interaction products of photons and charges (electrons, ions). These are typically achieved using conventional detectors/spectrometers and pump-probe schemes for measuring ultrafast processes, while the description of the interaction process and the interpretation of the experimental results is based on semi-classical approaches. 

Intense laser--atom interaction is one of the most fundamental processes in SLP. For this reason and for introducing the terminology and definitions relevant to the present article, in the next section, we will discuss a few of the fundamentals of intense laser--atom interaction.

\subsection{Intense laser--atom interaction} 
\begin{figure*}
    \centering
    \includegraphics[width=0.9 \linewidth]{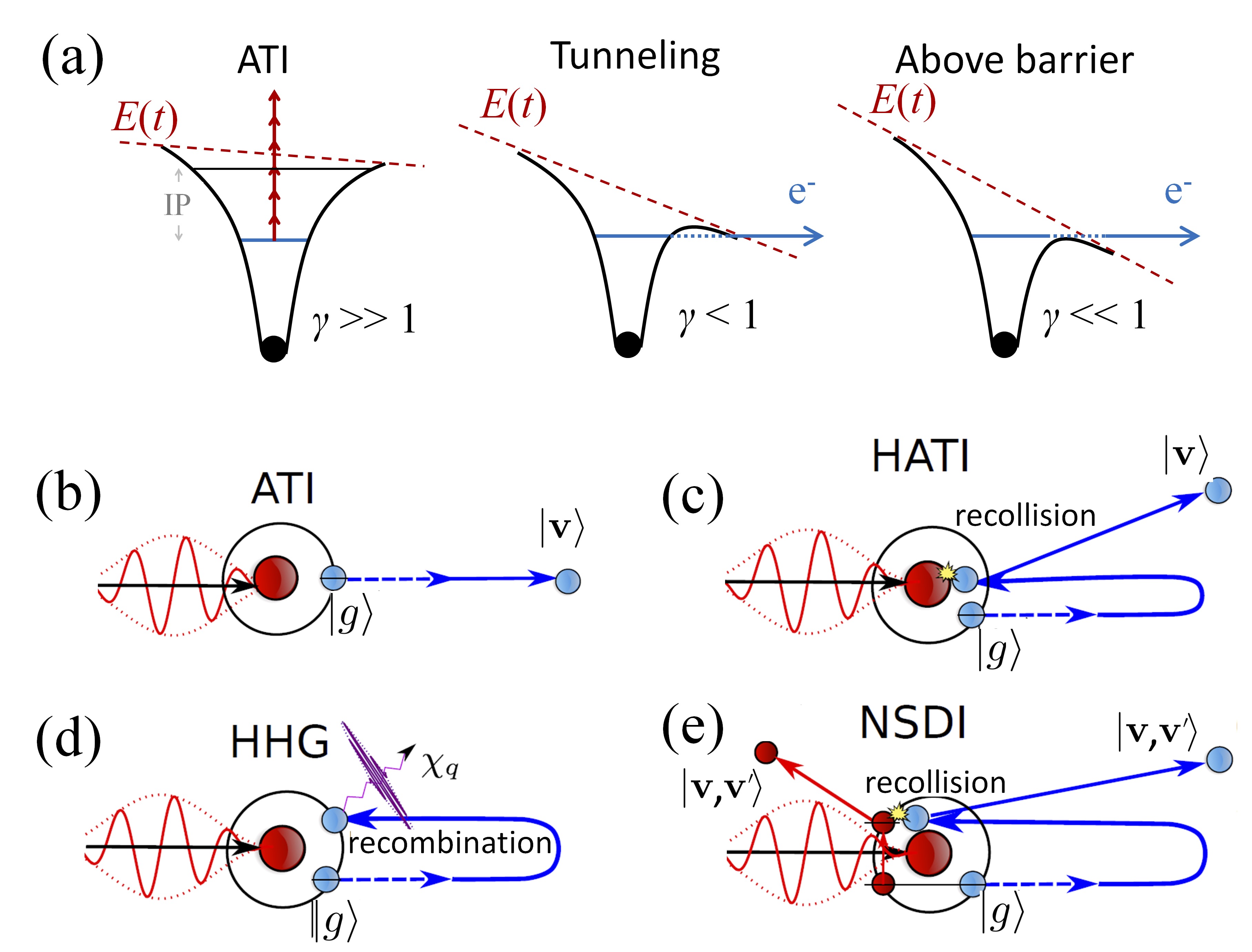}
    \caption{(a) A simplified drawing of intense laser--atom interaction. An IR laser pulse is focused into a target area which contains atoms (typically noble gases). The intensity of the laser pulse in the interaction area is typically $I_{L}>10^{14}$ W/cm$^2$. The atomic Coulomb potentials under the influence of the laser field $E(t)$ are shown for ATI, tunnel ionization, and above-barrier ionization cases. The multiphoton absorption is depicted by red vertical arrows. IP is the ionization potential and $\gamma$ is the Keldysh parameter. The electron tunnels out from the atomic potential bent by the laser field, then accelerates in the laser field from which it gains kinetic energy and then it may recollide elastically or inelastically with the parent ion. The process is repeated every half--cycle of the driving laser field and leads to the generation of electrons (via ATI and high--order ATI (HATI) processes), ions (via NSDI) and high harmonic photons. (b)--(e) Schematics of HHG, ATI, HATI and NSDI processes. The initial and final electronic states are given by $\ket{g}$ and $\ket{\mathbf{v}}$, respectively. Two \emph{classical} light fields are depicted, in red is the incident intense infrared laser field, while the harmonic emission is in purple.}
    \label{fig:Fig_Intense_laser_atom}
\end{figure*}

The main features of intense laser--atom interaction can be understood using theoretical tools developed over the past decades, starting with the seminal work by Keldysh in the 1960’s \cite{keldysh_ionization_1965, Perelomov1966, Reiss1980, Ammosov1986, Faisal_book_1987}. According to Keldysh theory, an electron can be freed from an atomic core via multiphoton, tunnel or above-the-barrier ionization. These regimes can be characterized by the Keldysh parameter $\gamma=\sqrt{I_p/(2U_{p})}$, where $I_p$ is the ionization potential of the atoms, $U_{p}=F_0^2/(4\omega_{L}^2) \approx 9.33 \cdot 10^{-14} \ [I_{L}(\text{W}/\text{cm}^2)\lambda_{L}^{2}(\mu \text{m})$] is the ponderomotive energy, i.e., the average oscillation energy of the electron in a laser field with amplitude $F_0$ and frequency $\omega_{L}$. The multiphoton ionization is dominating when $\gamma\gg 1$, while for $\gamma<1$ and $\gamma\ll 1$ the dominant processes are the tunnelling and the above barrier ionization, respectively (Fig.~\ref{fig:Fig_Intense_laser_atom}a). The majority of the investigations in HHG and attosecond science are performed in experimental conditions where $\gamma<1$ and the harmonic emission is phase matched. For this reason, here we omit from our discussion the case of $\gamma\gg1$. In the intensity regime where $\gamma<1$, the electron tunnels out from the atomic potential bent by the laser field $E(t)$, then accelerates in the laser field from which it gains kinetic energy and then it may recollide elastically or inelastically with the parent ion. The interaction has been successfully described by classical or semi–classical models, namely the three-step model \cite{kulander_dynamics_1993, corkum_plasma_1993, lewenstein-theory-1994, ABC19}, treating the electromagnetic field classically. This recollision process is repeated every half--cycle of the driving laser field and leads to the generation of electrons via ATI and high--order ATI (HATI) processes (Fig.~\ref{fig:Fig_Intense_laser_atom}b,c), high harmonic photons (Fig.~\ref{fig:Fig_Intense_laser_atom}d) and doubly charged ions via NSDI (Fig.~\ref{fig:Fig_Intense_laser_atom}e).  

\subsubsection{Above-threshold ionization (ATI) and high-order ATI}\label{Sec:ATI:Scl}
\hfill \break
The first strong-field effect that was conclusively demonstrated in experiment is ATI,
i.e., the ionization of a photoelectron through the absorption of more photons than it requires to overcome the ionization barrier. This can range from the so-called `multiphoton' regime at relatively low fields, where photon-absorption pictures can provide good models, to higher fields where perturbation theory completely breaks down, and quasi-static pictures (typically known as optical tunneling) provide a better description. One of the defining features of ATI is a ``comb'' in the photoelectron energy spectrum, where each peak is separated by one photon in energy, and which arises in the time domain as the interference of a sequence of photoelectron wavepackets produced with the periodicity of the driving laser.

The earliest observations of ATI~\cite{agostini_freefree_1979} required only a few photons, and could still be explained within lowest-order perturbation theory (LOPT). However, later results~\cite{chin_tunnel_1985} demonstrated a clear breakdown of perturbation theory, leading to the application of tunneling theories such Keldysh-Faisal-Reiss (KFR) theory~\cite{keldysh_ionization_1965, faisal_multiple_1973, Reiss1980} and, eventually, to the development of the widely applied strong-field approximation (SFA) \cite{lewenstein_rings_1995, becker_light_1995} and its semi-classical interpretation, which linked to the classical picture, the three-step model~\cite{corkum_plasma_1993, kulander_dynamics_1993}. For a complete review on the history and the current status of the SFA, see Ref.~\cite{ABC19}; see also \cite{armstrong-dialogue-2021} for a wider review of the theoretical methods of attoscience and strong-field physics. 

The three-step model for ATI provides a clear physical picture of the process, separated into two components: (i) direct ATI, shown in Fig.~\ref{fig:Fig_Intense_laser_atom}b, where an electron is released from a target via a strong laser field and propagates to the detector driven predominantly by the laser field and without significant further interactions with its parent ion, and  (ii) high-order ATI (HATI), shown in Fig.~\ref{fig:Fig_Intense_laser_atom}c, where the ionized electron undergoes a laser-driven elastic recollision with its parent.  The direct ATI process is associated with a relatively low photoelectron energy, with a classical cutoff of $2U_\mathrm{p}$. In contrast, the HATI process is associated with a much higher photoelectron energy, and it is typically present as a long plateau with a cutoff at $10U_\mathrm{p}$~\cite{paulus_plateau_1994}, which directly mimics the classical dynamics of the three-step model.  However, the recollision dynamics is generally much richer -- and much more quantum mechanical -- than the three-step model, and indeed the returning electron comes back as a high-energy wavepacket with a very short wavelength which can be used, among other things, to form images of the parent atomic or molecular ion, as is the case in laser-induced electron diffraction (LIED)~\cite{pullen-imaging-2015, wolter-ultrafast-2016}, as discussed below, or in photoelectron holography \cite{huismans_timeresolved_2011,figueirademorissonfaria_it_2020}.

\subsubsection{High harmonic generation}
\hfill \break
The recollision process in HATI between the tunneled photoelectron and its parent ion is one of the core dynamical features of strong-field physics, because -- in addition to the elastic scattering in HATI -- it can lead to a variety of outcomes, including inelastic scattering and further ionization of the parent. Most notably, however, it can also lead to the recombination of the photoelectron, which now carries significant kinetic energy, back into the electronic hole it left in the parent. That recombination results in the emission of a high-frequency burst of radiation, in the form of a sharp pulse of light that is typically as short as a few tens of attoseconds. These pulses normally form a periodic train of pulses whose frequency spectrum forms harmonics of the laser driver, which prompts this emission to be named HHG (Fig.~\ref{fig:Fig_Intense_laser_atom}d).

Informally, this emission is typically thought of as the emission of a single photon that carries away the kinetic energy of the photoelectron. One of the key goals of this work is to elucidate the degree to which this photon picture has a rigorous matching theoretical description within the QO realm. The emitted radiation, similarly to the photoelectron in HATI, normally forms a harmonic frequency comb with a long and flat plateau which terminates at a sharp cutoff that gives way to an exponential decay of the emission. The location of this cutoff is governed by the so-called ``cutoff law''~\cite{lewenstein-theory-1994, ABC19, pisanty-imaginary-2020}, $\hbar\Omega_\mathrm{cutoff}=3.17U_p+1.32I_p$, which includes the classical three-step model's prediction~\cite{corkum_plasma_1993, kulander_dynamics_1993} that the maximal recollision energy of the electron is $3.17U_p$, with the addition of the binding energy $I_p$, which is suitably amended by a quantum correction factor of 1.32~\cite{lewenstein-theory-1994, pisanty-imaginary-2020}. 

Experimentally, HHG was first observed at the end of the 1980s~\cite{ferray-multiple-1988, mcpherson-studies-1987}, and quickly developed into an attractive light source of attosecond pulses based on the realization that the emitted harmonics are phase-locked~\cite{salieres-study-1997}. As the subject has matured, both the microscopic~\cite{lewenstein-principles-2008} and macroscopic phase-matching~\cite{weissenbilder-how-2022, khokhlova-highly-2020, johnson-high-2018, popmintchev-extended-2008} aspects of HHG have come into sharper control, allowing the subject to expand into a wide variety of directions, from  studies of the conservation laws of the HHG emission when seen as a parametric nonlinear optical process~\cite{perry-high-1993, bertrand-ultrahigh-2011, FKD14, pisanty-spin-2014, zurch-strong-2012, hernandezgarcia-attosecond-2013, PRS19}
to multicolour tailoring of the waveform to optimize the harmonic cutoff~\cite{chipperfield-ideal-2009, haessler-optimization-2014, jin-waveforms-2014},
detailed tailoring of the polarization state of the harmonic emission~\cite{FKD14, KBH16, pisanty-spin-2014}, the harnessing of resonances in the continuum states of the target~\cite{ganeev-experimental-2012, strelkov-high-2014, khokhlova-polarization-2021, armstrong-dialogue-2021},
and the control over optical singularities in the emitted beam~\cite{zurch-strong-2012, hernandezgarcia-attosecond-2013, PRS19, RDB19},
among many others.

\subsubsection{Non-sequential double ionization}
\hfill \break
The above processes, HATI and HHG, elucidate the cases where a laser-driven recollision leads to an elastic recollision or recombination, respectively. However, if the recolliding photoelectron undergoes an inelastic recollision this may lead to further ionization via non-sequential double ionization (NSDI) or non-sequential multiple ionization (NSMI). The first evidence for NSDI and NSMI was found in \cite{lhuillier_multiply_1983}, where ionization yields of multiply ionized targets did not match 
the predictions corresponding to a sequential ionization mechanism.
The change in ionization rate from the non-sequential regime to the sequential regime, as laser intensity is increased, gave rise to the famous knee structure \cite{walker_precision_1994}. This non-sequential behaviour was ultimately attributed to a recollision mechanism \cite{moshammer_momentum_2000, weber_correlated_2000}, described as a three-step mechanism \cite{corkum_plasma_1993}, in direct analogy with HHG and HATI. The semi-classical processes associated with NSDI is depicted in Fig.~\ref{fig:Fig_Intense_laser_atom}e. One of the strongest indicators of recollision in NSDI was the high correlation between the two electrons \cite{weber_correlated_2000}. Recently, it has been shown in some cases, specifically for lower laser intensities, the orbital angular momentum of the two photoelectrons is entangled \cite{maxwell_entanglement_2022}.

\subsubsection{The strong field approximation}
\hfill \break
Strictly speaking, the interaction of matter with intense laser fields is described by the time-dependent Schrödinger equation (TDSE) that captures both the evolution of the wave functions and the time evolution of the physical observables. However, the solution of the TDSE in all the degrees of freedom of the system is computationally
very demanding. Moreover, a physical interpretation of the numerical results is highly nontrivial. Considering this, approximate methods are more than welcome. Such a method is the Strong Field Approximation (SFA) \cite{lewenstein-theory-1994}, which has consistently been shown over the years to be the workhorse tool for describing intense laser-matter interactions. Because the method has been extensively studied and reported over the years (c.f.~\cite{ABC19}), here, we will mention only the very fundamentals in case of intense laser-atom interactions.

The non-relativistic TDSE of an atom interacting with a single-mode long-wavelength (IR) linearly-polarized laser field ${\vb{E}_{\text{cl}}}(t)$ of frequency $\omega_{L}$ reads, 

\begin{equation} \label{eq:TDSE}
\begin{aligned}
i\hbar \pdv{\ket{\Psi(t)}}{t}
        = \hat{H}
            \ket{\Psi(t)},
 \end{aligned}
\end{equation}
where the Hamiltonian, $\hat{H} = \hat{H}_a + \hat{H}_{\text{int}}$, describes the laser-target system in the single-active-electron (SAE) approximation. Here, $\hat{H}_a$ is the atomic Hamiltionian and $\hat{H}_{\text{int}}=e{\vb{E}_{\text{cl}}}(t) \cdot {\bf \hat{r}}$ is the interacting Hamiltonian under the dipole approximation which describes the coupling of the atom, via the dipole operator $\hat{\vb{d}} = e \hat{\vb{r}}$, to a classical field $\vb{E}_{\text{cl}}(t)$.

\begin{figure}
    \centering
    \includegraphics[width=0.85 \columnwidth]{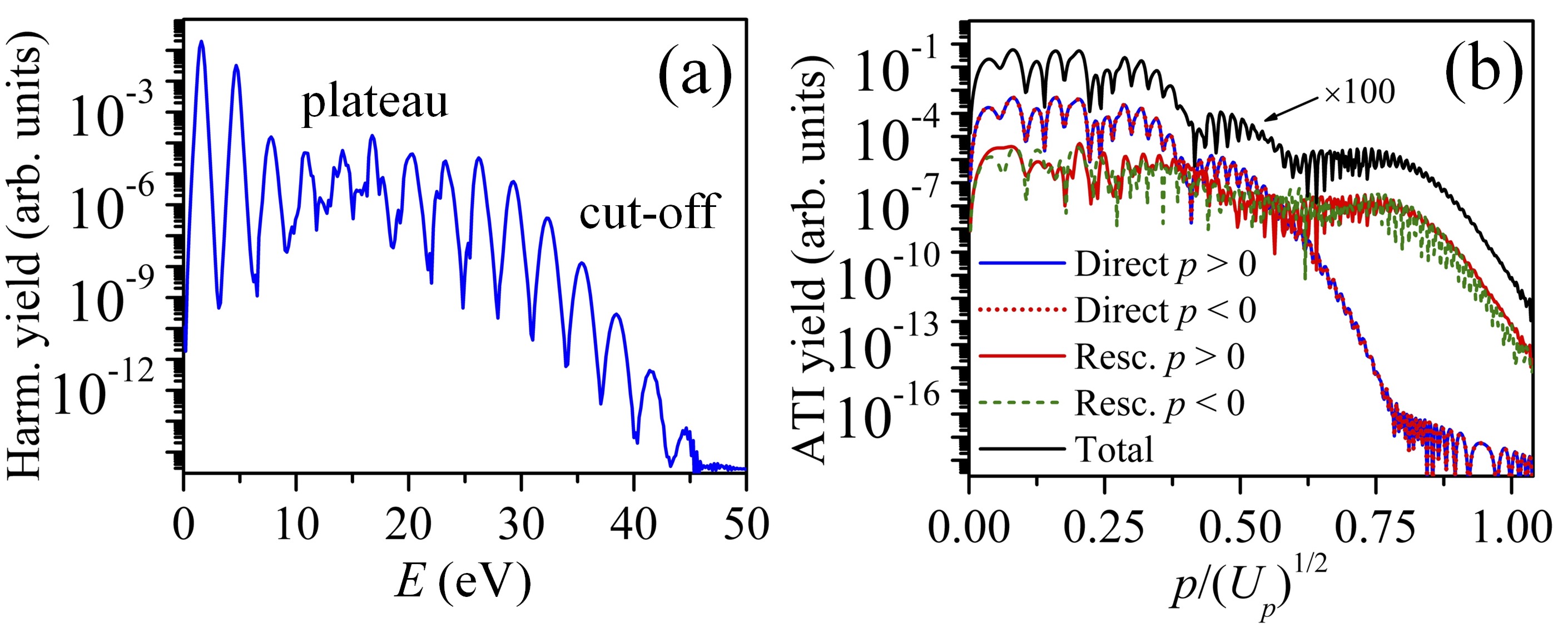}
    \caption{HHG (a) and ATI (b) spectra calculated when Xenon atoms interact with a linearly polarized laser pulse of intensity $8 \times 10^{13}$ W/cm$^2$ and $\sim\!30$ fs duration. In (b) the photoelectron spectrum corresponding to direct, rescattered photoelectrons with positive ($p > 0$) and negative ($p < 0$) momenta are shown with different colors. The ATI spectrum shown with black solid curve, which includes the contribution of the direct and rescattered electrons, has been shifted by a factor of 100 for visualization reasons. The figure has been reproduced from Ref.~\cite{SRM2023}.}
    \label{Fig:HHGATI}
\end{figure}

In the region of $\gamma < 1$, the effects of atomic effective potential on the dynamics of electrons in the continuum are assumed to be small, and they can be treated using perturbation theory. This suggests the following formulation of the ‘standard SFA’:
 (i) The electronic dynamics occurs in the subspace spanned by the ground state $|\text{g}\rangle$ and the scattering states $|\bf{v}\rangle$ (where ${\bf v}$ is the electron momentum). That is, excited bound states are neglected. 
(ii) The amplitude of the ground state is considered to be $\approx 1$ i.e., the ground state of the atom is not depleted. (iii) In the continuum, the electronic states are taken from the basis of exact scattering states, which are eigenstates of $\hat{H}_a \ket{{\bf v}}=\frac{{\bf v}^2}{2m} \ket{{\bf v}}$ with fixed momentum ${\bf v}$. However, the dominance of the strong laser field allows the perturbative expansion of the continuum states about the binding field, where to zeroth order, the electron can be considered as a free particle moving in the laser field without the influence of the atomic potential.
After making these approximations, the wave function can be written as $\ket{\Psi(t)}=e^{iI_pt}(\ket{\text{g}}+\int \dd^3{\bf v} \ b({\bf v}, t) \ket{{\bf v}})$ and the TDSE, Eq.~\eqref{eq:TDSE}, can be solved exactly with the $b({\bf v}, t)$ to be in a closed form. Thus, the ion, electron and HHG spectra can be calculated (using the saddle point method) by the ionization amplitudes $b({\bf v}, t)$ and the dipole matrix element $\bra{\Psi(t)} \hat{\vb{r}} \ket{\Psi(t)}$, respectively. Fig.~\ref{Fig:HHGATI} shows a representative example of the HHG and ATI spectra that have been calculated (using the \textsc{Qprop} software \cite{Qprop}) when Xenon atoms interact with a linearly polarized laser pulse of intensity $8 \times 10^{13}$ W/cm$^2$ and $\sim\!30$ fs duration.

\section{Intense laser--atom interaction: Fully quantized approach} \label{Sec:QED_Atoms}

The fully quantized description of intense laser--atom interaction and the conditioning approach that has been used for the generation of non-classical and entangled states has been extensively discussed in recent publications \cite{LCP21,RLP22,SRL22,SRM2023,RSM22}. Here, we summarize the main findings, emphasizing the results associated with potential applications in quantum information science.  

\begin{figure*}
    \centering
    \includegraphics[width=1.0 \linewidth]{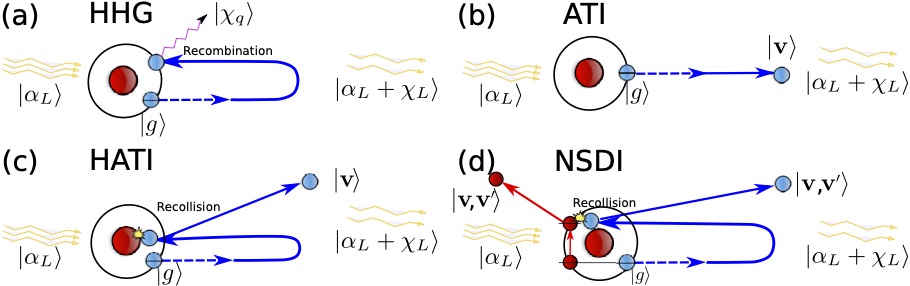}
    \caption{A schematic of the fully quantized description of intense laser--atoms interaction. From (a) to (b), the HHG, ATI, HATI and NSDI. Reproduced from Ref.~\cite{SRM2023}.}
    \label{fig:SF_QED}
\end{figure*}

The fully quantized description of intense laser--atom interaction is shown schematically  in Fig.~\ref{fig:SF_QED}, where 
$\ket{\alpha_{L}}$ and $\ket{\alpha_{L}+\chi_L}$ are the states of the driving field mode before and after interaction, respectively, $\ket{\chi_q}$ are the states of the generated high-order harmonic modes, and $\ket{g}$ and $\ket{\mathbf{v}}$ are the initial and final electronic states, respectively. It is noted that the electrons in NSDI may be entangled \cite{maxwell_entanglement_2022}, and therefore are written as a two-particle state. The quantum optical degrees of freedom may also be entangled with each other and the electronic states, however, for clearer labelling we have written these separately.

\subsection{Quantum electrodynamics of intense laser--atom interactions}

We start by considering a situation where only a single electron of the atom participates in the dynamics, and is initially in the ground state of the system, which we denote by $\ket{\text{g}}$. We characterize the laser field with a coherent state of amplitude $\alpha_L$, populating the mode of frequency $\omega_L \in\text{IR}$, while all the other modes remain in a vacuum state, i.e., $\ket{\Phi_i} = \ket{\alpha}\bigotimes^{\text{N}_{\text{c}}}_{q=2} \ket{0_q}$. For the sake of simplicity, we consider a discrete set of modes of frequency $\omega_q = q \omega_L$, $q \in \mathbbm{N}$ and introduce a cutoff frequency $N_{\text{c}}\omega_L$, beyond which no harmonics are considered. We set this frequency to be higher or equal to the cutoff frequency of the HHG spectrum. With all the above, we can write the initial state of the system prior to the dynamics as $\ket{\Psi(t_0)} = \ket{\text{g}}\otimes \ket{\alpha_L}\bigotimes^{\text{N}_{\text{c}}}_{q=2} \ket{0_q}$.

\begin{figure}
    \centering
    \includegraphics[width=1.0 \textwidth]{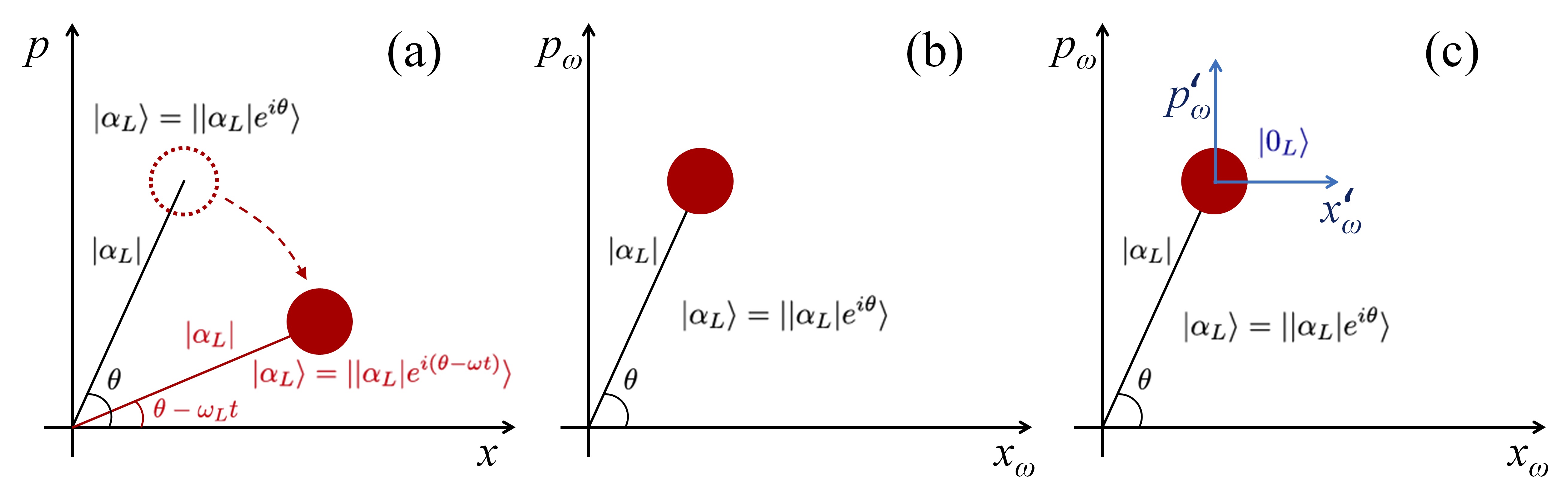}
    \caption{Pictorial representation showing how the unitary transformations affect our frame of reference in the quantum optical phase space. (a) Initially, if no transformations are applied and we consider only the dynamics due to $H_f$, the coherent state of the field rotates with frequency $\omega_L$ around the phase space origin. (b) When we work in the interaction picture with respect to $H_f$, our frame of reference oscillates with frequency $\omega_L$, since the time-dependence due to this Hamiltonian goes to the operators, and the quantum optical state gets fixed in the phase space. (c) Finally, we perform a translation of our frame of reference, such that its origin is now located on top of the coherent state. Thus, in this new frame, the coherent state is given in terms of the vacuum state $\ket{0_L}$.}
    \label{Figs:QO:transf}
\end{figure}

The Hamiltonian characterizing the dynamics between the laser and a single electron of the corresponding atom, is given by $\hat{H} = \hat{H}_a + \hat{H}_f + \hat{H}_{\text{int}}$, where $\hat{H}_a$ is the atomic Hamiltonian, $\hat{H}_f$ is the Hamiltonian of the electromagnetic field (as given in Sec.~\ref{Section:fund:QO}), and $\hat{H}_{\text{int}}$ describes the interaction between light and matter. Under the length-gauge and dipole approximation, the latter can be written as $\hat{H}_{\text{int}} = \hat{\vb{d}}\cdot \hat{\vb{E}}$, with $\hat{\vb{d}}$ the dipole moment operator and $\hat{\vb{E}} =  i \vb{g}(\omega_L) \sum_{q=1}^{N_c} (a_q - a_q^\dagger) $ the electric field operator. In order to simplify the description of these dynamics, we move to a more convenient frame by applying a set of unitary transformations, whose effect on the quantum optical state in phase space is pictorially presented in Fig.~\ref{Figs:QO:transf}. First, we move to the interaction picture with respect to $H_{f}$, which makes our frame of reference to rotate with the frequency of the field (see Fig.~\ref{Figs:QO:transf}b), and therefore introduces a time-dependence in the electric field operator, i.e. $\hat{\vb{E}} \to \hat{\vb{E}}(t)$ with $\hat{a}_q \to \hat{a}_q e^{-i\omega_q t}$. Second, we displace our frame of reference in phase-space by $\alpha_L$, such that the initial IR coherent state $\ket{\alpha_L}$ is in the origin of the shifted frame. To account for this shift in the Hamiltonian, the electric field splits into two components, i.e. $\hat{\vb{E}}(t) \to \vb{E}_{\text{cl}}(t) + \hat{\vb{E}}(t)$, where the first is the mean value of the electric field in the initial state $\lvert \Phi_i(t_0)\rangle = \ket{\alpha_L}\bigotimes_{q=2}^{N_{\text{c}}} \ket{0_q}$, i.e. $\vb{E}_{\text{cl}}(t) = \Tr[\hat{\vb{E}}(t) \dyad{\Phi_i(t_0)}]$, and the second term provides the quantum fluctuations around the mean value. Finally, we work in the interaction picture with respect to the \emph{semiclassical} Hamiltonian $\hat{H}_{\text{sc}}(t) = \hat{H}_a + \hat{\vb{d}}\cdot \vb{E}_{\text{cl}}(t)$, which makes the dipole operator time-dependent, i.e., $\hat{\vb{d}}\to \hat{\vb{d}}(t)$. With this last transformation, we encapsulate all the dynamics found in the semiclassical analysis \cite{lewenstein-theory-1994,ABC19}, into the time-dependent dipole operator, and we are thus left with a Hamiltonian $\hat{H}^\prime(t) = \hat{\vb{d}}(t) \cdot \hat{\vb{E}}(t)$ coupling the electron dipole moment to the electric field operator. The corresponding Schrödinger equation is given by
\begin{equation}\label{Eq:Sch:transf}
    i\hbar \dv{\ket{\psi(t)}}{t} 
        = \hat{\vb{d}}(t) \cdot \hat{\vb{E}}(t)
            \ket{\psi(t)},
\end{equation}
and the initial condition is now given by $\ket{\psi(t_0)} = \ket{\text{g}}\bigotimes^{\text{N}_{\text{c}}}_{q=1} \ket{0_q}$, where $q=1$ corresponds to the driving IR frequency. This differential equation, constitutes the basis of the upcoming analysis.

\subsubsection{Conditioning on HHG using single--color driving fields: Generation of optical ``cat'' and entangled states from XUV to IR}
\label{conditioningHHG}
\hfill\break

In order to characterize the quantum optical state after the HHG processes, where the electron is found on its ground state, we project Eq.~\eqref{Eq:Sch:transf} onto $\ket{\text{g}}$, and denote the state of light as $\ket{\Phi(t)} = \braket{\text{g}}{\psi(t)}$. We further introduce the SFA version of the identity, $\mathbbm{1} = \dyad{\text{g}} + \int \dd \vb{v} \dyad{\vb{v}}$, where the contribution of the excited bound states have been neglected since they barely participate in the dynamics, in virtue of the strong-field approximations \cite{lewenstein-theory-1994}. However, further neglecting the contribution of the electronic continuum states $\ket{\vb{v}}$, since the continuum amplitude is much smaller than the ground state amplitude, we get that the dynamics of $\ket{\Phi(t)}$ can be described by
\begin{equation}\label{Eq:Sch:HHG}
    i\hbar \dv{\ket{\Phi(t)}}{t}
        = \expval{\vb{d}(t)}\cdot \hat{\vb{E}}(t)
            \ket{\Phi(t)},
\end{equation}
where $\expval{\vb{d}(t)} = \mel{\text{g}}{\vb{d}(t)}{\text{g}}$ is the time-dependent expectation value of the dipole moment with respect to the ground state.

Since the commutator of the new interaction Hamiltonian in \eqref{Eq:Sch:HHG} at different times is just a complex number, the different field modes will not couple when solving the Schrödinger equation in \eqref{Eq:Sch:HHG}. Thus, each of the field modes undergoes an independent evolution, and a \emph{classical} oscillating current is coupled to each of the modes. As discussed in Sec.~\ref{Sec:Coherent:States}, the unitary evolution corresponding to a Hamiltonian of this form can be written in terms of the displacement operator, inducing a shift in the phase-space of the corresponding quantum optical mode. Hence, we find that the quantum optical state of the system after these dynamics is given by
\begin{equation}\label{Eq:Sol:HHG}
    \ket{\Phi(t)} 
        = e^{i\varphi(t,t_0)}\bigotimes^{N_{\text{c}}}_{q=1}
            \hat{D}\big(\chi_q(t,t_0)\big)
                \ket{0_q}.
\end{equation}

As an output, we get a coherent displacement in the quantum optical field modes of a quantity $\chi_q(t,t_0)$, given by
\begin{equation}
    \chi_q(t,t_0) 
        = -\dfrac{1}{\hbar}
            \int^t_{t_0} \dd t' \boldsymbol{g}(\omega_L)
                \cdot \expval{\vb{d}(t)} e^{i\omega_q t},
\end{equation}
which corresponds to the Fourier transform of the time-dependent dipole moment. The absolute value of this quantity, when considering the limit $t\to \infty$, leads to the known features of the HHG spectrum \cite{SRM2023}. The approximation leading to \eqref{Eq:Sch:HHG} has the underlying assumption of vanishing correlations in the dipole moment operator \cite{Sta22,sundaram_high-order_1990}. We note that, so far, we have been working under single-atom dynamics, and therefore the generated shift $\chi_q(t,t_0)$ is very small. Nevertheless, this solution can be extended to a more realistic scenario, where we have $N_{\text{ph}}$ atoms participating in the dynamics in a phase-matched way, by multiplying the shift with the number of phase-matched atoms $N_{\text{ph}}$, i.e., 
\begin{equation}\label{Eq:NrAtoms}
    \chi_q(t,t_0) \to N_{\text{ph}}\chi_q(t,t_0).
\end{equation}

The solution we have obtained in Eq.~\eqref{Eq:Sol:HHG} is given by the tensor product of the coherent states $\{\ket{\chi_q(t,t_0)}\}_q$, and is therefore what we referred to in Sec.~\ref{Section:2} as a  \emph{classical} state. Nevertheless, by introducing conditioning measurements on the harmonic field modes, we could potentially generate non-classical states of light in the form of coherent state superpositions and entangled coherent states between different frequency modes. Such conditioning operations are possible because, as a consequence of the light-matter interaction, each of the field modes gets excited from the same dipole moment, making the respective shifts $\chi_q(t,t_0)$ correlated. Thus, the mode that is actually excited during the HHG process is a wavepacket mode taking into account these correlations \cite{LCP21,SRL22,Sta22,SRM2023}, and that can be described with a set of number states $\{\ket{\Tilde{n}}\}$. Here, the state $\ket{\Tilde{0}}$ describes the case where no harmonic radiation is generated, and therefore corresponds to the initial quantum optical state. On the contrary, $\ket{\Tilde{n}}$ with $\Tilde{n}\neq 0$, describes the case where harmonic radiation has been generated. This allows us to define a set of positive operators $\{\Pi_{\Tilde{0}}, \Pi_{\Tilde{n}\neq \Tilde{0}}\}$, describing whether harmonic radiation is generated or not, or equivalently whether the wavepacket which takes into account the correlations between the field modes has been excited or not. Specifically, the element $\Pi_{\Tilde{0}} = \dyad{\Tilde{0}}$ projects onto the subspace where no excitations are found, while $\Pi_{\Tilde{n}\neq \Tilde{0}} = \sum_{\Tilde{n}\neq \Tilde{0}}  \dyad{\Tilde{n}} = 1-\dyad{\Tilde{0}}$ onto the subspace where HHG excitations are found. The \emph{conditioning on HHG} operation corresponds to this second case, and can be witnessed by the detection of harmonic radiation. Applying this operation onto the state given in Eq.~\eqref{Eq:Sol:HHG} leads to
\begin{equation}\label{Eq:HHG:cond}
    \ket{\Phi_{\text{HHG}}(t)}
        = \ket{\alpha_L + \chi_1(t,t_0)}
            \bigotimes^{N_{\text{c}}}_{q=2}
                \ket{\chi_q(t,t_0)}
            -\xi_1(t,t_0) \ket{\alpha_L} \bigotimes^{N_{\text{c}}}_{q=2}
                \xi_q(t,t_0)\ket{0_q},
\end{equation}
where in this expression we have undone the displacement transformation $\hat{D}(\alpha_L)$ introduced in order to arrive to Eq.~\eqref{Eq:Sch:transf}, and we have defined $\xi_1(t,t_0) = \braket{\alpha}{\alpha + \chi_L(t,t_0)}$ and $\xi_q(t,t_0) = \braket{0}{ \chi_q(t,t_0)}$.

\begin{figure}
    \centering
    \includegraphics[width =0.9 \textwidth]{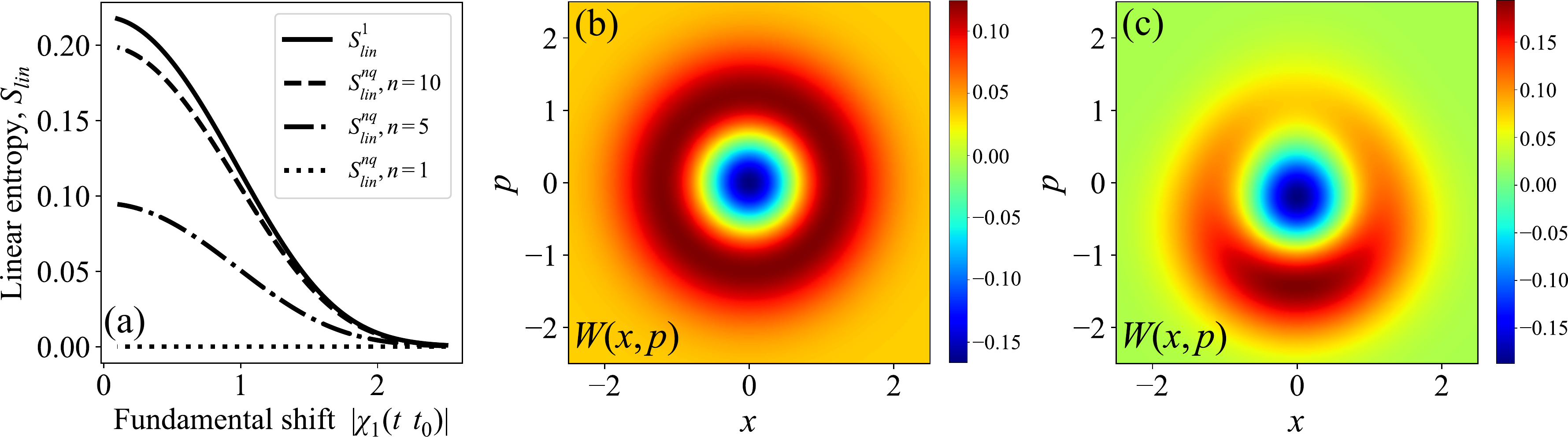}
    \caption{In (a) we show the behavior of the linear entropy as a function of the shift in the fundamental mode. The linear entropy is defined as $S_{lin}(\rho) = 1- \Tr(\rho^2)$, where $\rho$ is the partial trace of the entangled state in Eq.~\eqref{Eq:HHG:cond} with respect to some of the harmonic modes. This quantity can be used as a measure of entanglement between coherent states \cite{agarwal_quantitative_2005,berrada_beam_2013}. Specifically, $S^1_{lin}$ (solid curve) shows the entanglement between the fundamental mode and all the other harmonic modes, and $S_{lin}^{nq}$ shows the entanglement between $n$ harmonic modes and the rest (among which we include the fundamental). This figure has been adapted from Ref.~\cite{SRL22}. (b) and (c) show the calculated Wigner functions of the quantum state presented in Eq.~\eqref{Eq:HHG:Cond:IR} for two different shifts, namely $\abs{\chi_1(t,t_0)} = 0.01$ and $\abs{\chi_1(t,t_0)} = 0.5$, respectively.}
    \label{Fig:Ent:Wigner:HHG}
\end{figure}

The state presented in \eqref{Eq:HHG:cond} is a massively entangled state between the field modes. And we highlight \emph{massive} as the number of modes excited after HHG processes can easily surpass the limit $N_{\text{c}} > 10$. However, the entanglement properties of this state crucially depend on how big the shifts $\chi_q(t,t_0)$ are \cite{SRL22} (see Fig.~\ref{Fig:Ent:Wigner:HHG}a). Specifically, when these become too large, the quantities $\xi_q(t,t_0)$ become very small and the obtained state becomes separable. This can be seen from Eq.~\eqref{Eq:HHG:cond} when considering the limit $\xi_q(t,t_0)\to 0$, which leads to a separable state of the form $\ket{\Phi_{\text{HHG}}(t)}= \ket{\alpha_L + \chi_1(t,t_0)}\bigotimes^{N_{\text{c}}}_{q=2}\ket{\chi_q(t,t_0)}$. We note that the values of the shifts depend on the harmonic yield, and can be increased or decreased by, for instance, enlarging or reducing the number of atoms participating in the HHG process by varying the gas density~\cite{RLP22}.

\begin{figure}
    \centering
    \includegraphics[width=0.85 \textwidth]{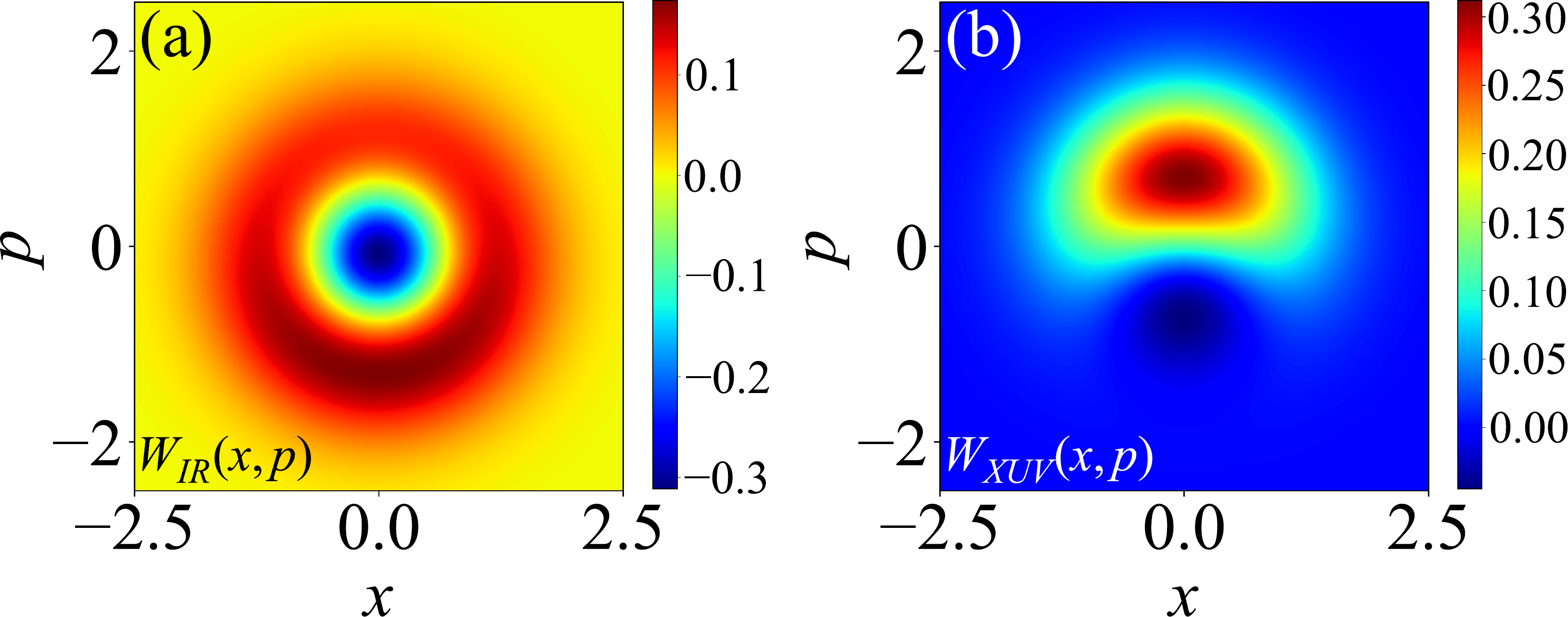}
    \caption{(a) and (b) show the calculated Wigner functions corresponding, respectively, to a coherent state superposition of the fundamental mode (typically of wavelength $\lambda_{\text{IR}} = 800$ nm), and to a coherent state superposition for the $q$th harmonic mode. For the calculation, the harmonic cutoff has been set to the 11th harmonic (corresponding to a wavelength of $\lambda_{\text{XUV}} = 72.7$ nm), $\chi_1 = -0.2$, such that $\chi_{q} \approx 0.03$. The opposite shift in the imaginary part reflects the correlation between the field modes. The figure has been reproduced from Ref.~\cite{SRL22}.}
    \label{Fig:XUVIR_CAT}
\end{figure}

As mentioned before, this scheme can be used for generating coherent state superpositions in the driving field, leading to high-photon number non-classical states of light. Specifically, projecting the state in Eq.~\eqref{Eq:HHG:cond} onto the state in which the harmonics are found, i.e. $\bigotimes_{q=2}^{N_{\text{c}}} \ket{\chi_q(t,t_0)}$, we get
\begin{equation}
\label{Eq:HHG:Cond:IR}
    \ket{\Phi^{(\text{IR})}_{\text{HHG}}(t)}
        = \ket{\alpha_L + \chi_1(t,t_0)}
            -\xi_1(t,t_0) \abs{\xi_{\text{HH}}(t,t_0)}^2 \ket{\alpha_L},
\end{equation}
where $\xi_{\text{HH}}(t,t_0) = \prod_{q=2}^{N_{\text{c}}} \xi_q(t,t_0)$. The value of $\chi_1(t,t_0)$ determines the distinguishability of the coherent states in the superposition, and therefore leads to the generation of ``kitten-like'' or ``cat-like'' states, as shown in Figs.~\ref{Fig:Ent:Wigner:HHG}b and c. On the other hand, when $\chi_1(t,t_0)$ becomes excessively large, the second term in the superposition shown in \eqref{Eq:HHG:Cond:IR} vanishes, and we are left with the coherent state $\ket{\alpha_L + \chi_1(t,t_0)}$. Note that we can equivalently perform the same operations onto an XUV mode, that is, to measure all the field modes (including the fundamental), except one of the harmonic ones. This allows us to obtain coherent state superpositions for frequencies belonging to the XUV region \cite{SRL22}, which are given for the mode with frequency $\omega_q = q \omega_L$ by
\begin{equation}\label{Eq:HHG:cond:XUV}
    \ket{\Phi_{\text{HHG}}^{(q)}(t)}
        = \ket{\chi_q(t,t_0)}
        -\xi_q(t,t_0) \abs{\bar{\xi}(q,t,t_0)}^2
        \ket{0},
\end{equation}
where we have defined $\bar{\xi}(q,t,t_0) = \prod_{q'\neq q} \xi_{q'}(t,t_0)$. In Fig.~\ref{Fig:XUVIR_CAT}, an example of the Wigner function obtained for Eq.~\eqref{Eq:HHG:Cond:IR} and Eq.~\eqref{Eq:HHG:cond:XUV}, is shown respectively in a and b, where the opposite shift in phase space reflects the correlation between the field modes \cite{SRL22}.

\subsubsection{Conditioning on HHG using two--color driving fields: Generation of optical ``cat'' and entanglement between the driving field modes}
\label{conditioningHHG_2colors}
\hfill \break

Using this conditioning technique, together with more elaborated HHG schemes, one can generate more entangled states using frequencies belonging to the visible and infrared range. For instance, we can excite the atom by using a two-color driving field, as shown in Ref.~\cite{SRL22}. In these experiments \cite{kim_highly_2005,mauritsson_attosecond_2006,FKD14}, two-color laser fields of frequencies $\omega_1$ and $\omega_2$, belonging to the far-infrared and visible spectral regions, that have either parallel or orthogonal polarizations, are combined to drive the atomic system. In this case, the initial state of the field is $\ket{\alpha_{\omega_1}}\ket{\alpha_{\omega_2}}$, while the field state after the HHG process is $\ket{\alpha_{\omega_{1}} + \chi_{\omega_{1}}}\ket{\alpha_{\omega_{2}} + \chi_{\omega_{2}}} \bigotimes^{N_{\text{c}}}_{q} \ket{\chi_q}$. After performing the \emph{conditioning on HHG} operation \cite{SRL22}, the final quantum optical state can be written as
\begin{equation}\label{Eq:two_color}
    \ket{\Phi^{(\omega_1-\omega_2)}_{\text{HHG}}(t)}
        = \ket{\alpha_{\omega_{1}} + \chi_{\omega_{1}}}\ket{\alpha_{\omega_{2}} + \chi_{\omega_{2}}}
            -\xi_{(\omega_1,\omega_2)} \ket{\alpha_{\omega_{1}}}\ket{\alpha_{\omega_{2}}},
\end{equation}
where $\chi_{\omega_{1}}$, $\chi_{\omega_{2}}$ is the depletion of the two--color driving field modes. The factor $\xi_{(\omega_1,\omega_2)}$, is a complex number that depends on the phase and amplitude of the initial coherent states and of the generated shifts. 

\subsubsection{Conditioning on ATI and generation of optical ``cat'' states}
\label{conditioningATI}
\hfill \break

In ATI processes, the electron is found in the continuum after the end of the pulse (see Fig.~\ref{fig:Fig_Intense_laser_atom}). Therefore, with the aim of characterizing the quantum optical state after ATI processes, we project the TDSE in Eq.~\eqref{Eq:Sch:transf} with respect to a continuum state $\ket{\vb{v}}$. When doing so, and after introducing the SFA version of the identity, we find that considering only direct ATI events, i.e. at zeroth order level with respect to the rescattering matrix elements, we can write the TDSE as \cite{SRM2023}
\begin{equation}\label{Eq:Direct:ATI}
    i\hbar \dv{\ket{\Phi({\vb{v},t)}}{t}}{t}
        = \hat{\vb{E}}(t)\cdot \mel{\vb{v}}{\vb{d}(t)}{\text{g}} \ket{\Phi(t)}
        +\hat{\vb{E}}(t)\cdot \Delta \vb{r}
            \ket{\Phi(\vb{v},t)},
\end{equation}
where we have defined $\ket{\Phi(\vb{v},t)} = \braket{\vb{v}}{\psi(t)}$, and where $\Delta \vb{r}(\vb{v},t)$ is the total displacement performed by the electron during its propagation in the continuum, which is given by
\begin{equation}
    \Delta \vb{r}(t,\vb{v})
        = \dfrac{e}{m}\int^t_{t_0}\dd t'
            \bigg[
                m\vb{v} - \dfrac{e}{c}\vb{A}(t)
                + \dfrac{e}{c}\vb{A}(t'),
            \bigg]
\end{equation}
with $\vb{A}(t)$ the classical vector potential, defined as $\vb{E}_{\text{cl}}(t) = - \pdv*{\vb{A}(t)}{t}$.

The differential equation presented in \eqref{Eq:Direct:ATI}, is composed by two terms. The first, introduces the transition matrix element between the ground state and the continuum state, which characterizes the backaction of the ionization process over the field modes. On the other hand, the second term depends on the electronic displacement $\Delta \vb{r}(\vb{v},t)$, and therefore takes into account the backaction of the electronic motion in the continuum over the field modes. A solution to this differential equation, having in mind that the electron is initially found in the ground state, can be written as \cite{SRM2023}
\begin{equation}\label{Eq:Sol:ATI}
    \ket{\Phi(\vb{v},t)}
        = -\dfrac{i}{\hbar}
            \int^t_{t_0} \dd t'
                e^{i\varphi(t,t',\vb{v})}
                \prod_q
                \hat{D}\big(\delta_q(t,t',\vb{v})\big)
                \hat{\vb{E}}(t')
                \cdot \mel{\vb{v}}{\hat{\vb{d}}(t')}{\text{g}}
                \ket{\Phi(t')},
\end{equation}
where $\ket{\Phi(t')}$ is given as in Eq.~\eqref{Eq:Sol:HHG}, and we have defined
\begin{equation}
    \delta_q(t,t',\vb{v})
        = -\dfrac{1}{\hbar}
            \int^t_{t'} \dd \tau \ \vb{g}(\omega_q)
            \cdot \Delta \vb{r}(\tau,\vb{v}) e^{i\omega_q \tau},
\end{equation}
which is the Fourier transform of the electron's displacement. Thus, from the evolution given by Eq.~\eqref{Eq:Sol:ATI}, we see that the quantum optical state of the system gets displaced first by a quantity $\chi_q(t,t_0)$, as a consequence of the electronic oscillation in the ground state \cite{madsen_strong-field_2021}, leading to $\ket{\Phi(t^\prime)}$. At time $t'$ an ionization process takes place, which is highlighted in the analytical expression by the presence of the transition matrix element $\mel{\vb{v}}{\hat{\vb{d}}(t)}{\text{g}}$, where the electron gets promoted from the initial ground $\ket{\text{g}}$ to a continuum state $\ket{\vb{v}}$. This event, induces a change on the final state of the field happening at $t'$, and that is due to the coupling of this matrix element to the electric field operator. Finally, the electron propagates in the continuum driven by the field, as an oscillating and drifting charge, and therefore behaves as a classical charge current. This leads to a shift in the quantum optical state of the field given by $\delta_q(t,t',\vb{v})$ for the $q$th mode, and that depends on the final momentum $\vb{v}$ with which the electron is found. We note that these dynamics are for the case of direct ionization, with a kinetic energy of less than $2U_p$. However, the electron can still undergo rescattering processes, which define the high-energy part of the photoelectron spectrum (see Sec.~\ref{Sec:ATI:Scl}). These extra dynamics, where the electrons can end up with values of the kinetic energy up to $10U_p$, get reflected on the final quantum optical state by an extra coupling to the electromagnetic field happening when the rescattering takes place (see Ref.~\cite{SRM2023} for more details).

\begin{figure}
    \centering
    \includegraphics[width=0.8 \textwidth]{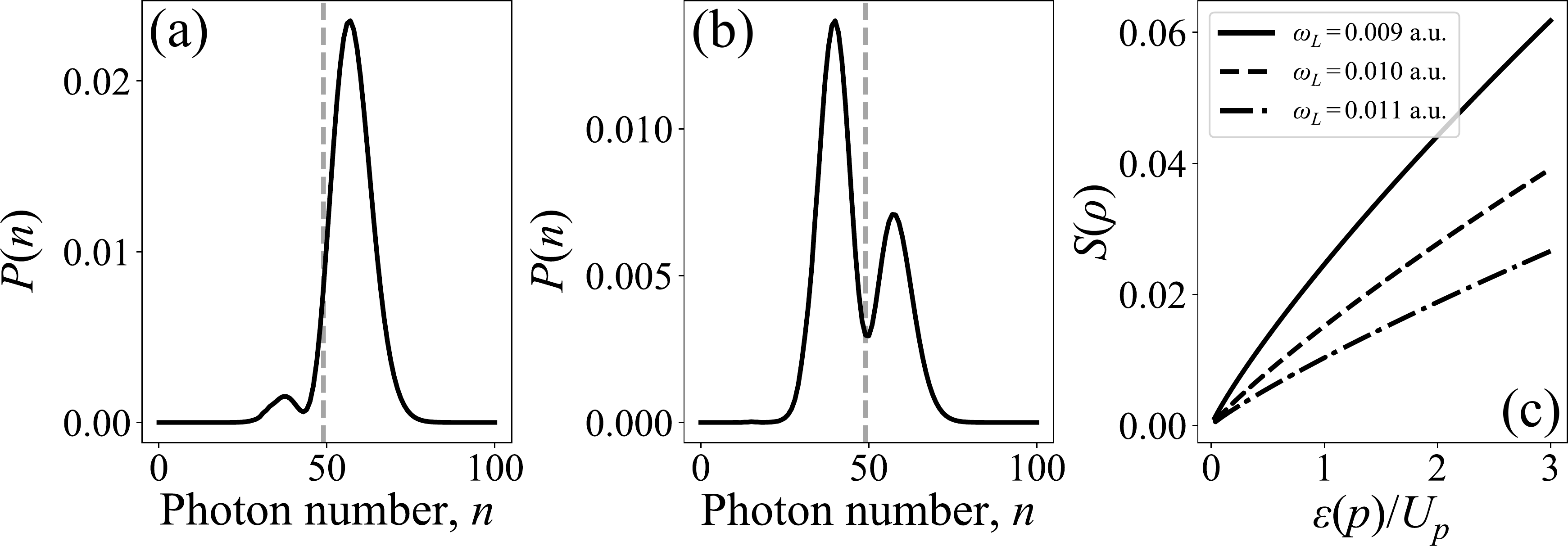}
    \caption{In (a) and (b), we show the photon number probability for two values of momentum, (a) $p = 0.32\sqrt{U_p}$ and (b) $p = -0.32\sqrt{U_p}$, where $p = m v - eA(t)/c$ is the canonical momentum. For representation purposes, the initial coherent state of the system has been set to $\abs{\alpha_L} = 7$. These figures have been extracted from Ref.~\cite{SRM2023}. In (c), we show the entropy of entanglement $S(\rho)$ as a function of the photoelectron energy $\mathcal{E}(\rho)/U_p$ for three different frequencies, $\omega_L = 0.009$ a.u. (solid curve), $\omega_L = 0.010$ a.u. (dashed curve) and $\omega_L = 0.011$ a.u. (dash-dotted curve). This figure has been extracted from Ref.~\cite{RSM22}.}
    \label{Fig:QED:ATI}
\end{figure}

One of the striking features of the quantum optical state after ATI in Eq.~\eqref{Eq:Sol:ATI}, is that the displacement obtained over the field modes, depends on the ionization time $t'$ of the electron, which naturally leads to a quantum superposition. This contrasts with what was found in HHG, where the final state of the system was given as a tensor product between different coherent states, and the superposition was generated by introducing the \emph{conditioning on HHG} operation. Furthermore, the quantum superpositions obtained for ATI, also depend on the final momentum of the electron $\vb{v}$. In this direction, it was found in Ref.~\cite{SRM2023} that the displacement $\delta(t,t',\vb{v})$, differed for positive and negative values of the momentum. Specifically, while showing a phase difference of $\pi$, the imaginary component of $\delta(t,t',\vb{v})$ has opposite signs for positive and negative momentum \cite{SRM2023,RSM22}. This is related to the fact that the sign of $\vb{v}$ determines the direction along which the electron is propagating, and therefore determines whether the generated displacement interferes in- or out-of-phase with the input driving field. This leads to either an enhancement or depletion of the fundamental mode, which can be observed by computing the photon number probability distribution $P(n)$, for $\ket{\Phi^{\text{IR}}(\vb{v},t)} = \braket{0_{\text{HH}}}{\Phi(\vb{v},t)}$, with $\ket{0_{\text{HH}}}$ a shorthand notation for $\bigotimes_{q>2} \ket{0_q}$. Here, we have assumed that the harmonic modes remain in the vacuum state $\ket{0_q}$, in order to account for the case where no harmonics are emitted \cite{SRM2023}. In Fig.~\ref{Fig:QED:ATI}a and b, we show $P(n)$ for two values of $\vb{v}$ that are equal in magnitude, but differ in sign. For representation purposes, the initial coherent state has been set to $|\alpha| = 7$, such that the vertical dashed line represents the initial mean photon number. We observe that, for positive momentum (Fig.~\ref{Fig:QED:ATI}a), the shift tends to enhance the mean photon number, while for negative momentum (Fig.~\ref{Fig:QED:ATI}b), it tends to lower it. Nevertheless, the influence of positive and negative momentum depends on the carrier-envelope phase, i.e. the phase difference between the envelope and the carrier wave of the pulse, such that a modification in $\pi$ of this phase, leads to the contrary effect. On the other hand, the double peak structure in both plots witness the fact that the final state is given as superposition between coherent states \cite{SRM2023,RSM22}.

This difference on the final quantum optical state of the field for positive and negative momentum, could be used for the generation of entangled states between light and matter \cite{RSM22}. In particular, when restricting to ATI processes, the total state of the system can be written as
\begin{equation}
    \ket{\psi(t)}
        = \int \dd \vb{v} \ket{\vb{v}}\ket{\Phi(\vb{v},t)},
\end{equation}
which has the form of an entangled state because different values of the electronic kinetic momentum lead to different effects on the field modes. Let us restrict now, for the sake of simplicity, to a 1D scenario, and to the case where only electrons of a well-defined energy are emitted. Then, the previous state could be written as $\ket{\psi(t)} = \ket{v}\ket{\Phi(v,t)} + \ket{-v}\ket{\Phi(-v,t)}$, and the amount of entanglement in it can be quantified in terms of the entropy of entanglement $S(\rho) = -\tr(\rho \log_2\rho)$ \cite{nielsen_quantum_2010,plenio_introduction_2007,van_loock_optical_2011}, with $\rho$ the reduced state of the system with respect to one of the parties. In Fig.~\ref{Fig:QED:ATI}c, $S(\rho)$ is shown as a function of the photoelectron's energy $\mathcal{E}(p)$ for a single-active electron, and for three different frequencies that belong to the MIR regime, when using a field with five optical cycles. Within this range, the values of $\abs{\delta_1(v,t',t)}$ become big enough when considering currently available field intensities for the corresponding frequency regime \cite{RSM22}. As observed, the amount of entanglement increases the bigger the value of the photoelectron's energy is, and gets enhanced the more time the electron spends oscillating in the field, i.e., the lower the frequency for fields with the same number of cycles. An important feature, required to obtain these entangled states, is that the process of emitting an electron with positive and negative momentum has to be equally probable. In other words, the photoelectron spectrum needs to be symmetric. Otherwise, one of the components will dominate over the other, which makes the final state of the system to be separable. This sort of symmetricity can be found when driving the system with multicycle laser pulses, that is, when the CEP effects become negligible, while in the case of laser pulses with few optical cycles this is not the case anymore \cite{protopapas_atomic_1997,milosevic_above-threshold_2006} (see Fig.~\ref{Fig:HHGATI}b). Finally, note that when the final state of the electron cannot be discerned, the quantum optical state is left in a mixed state of the form,
\begin{equation} \label{eq:cat_ATI_cont}
	\rho(t) = \int \dd \vb{v} \dyad{\Phi(\vb{v},t)}.
\end{equation}

\begin{figure}
    \centering
    \includegraphics[width=0.8 \textwidth]{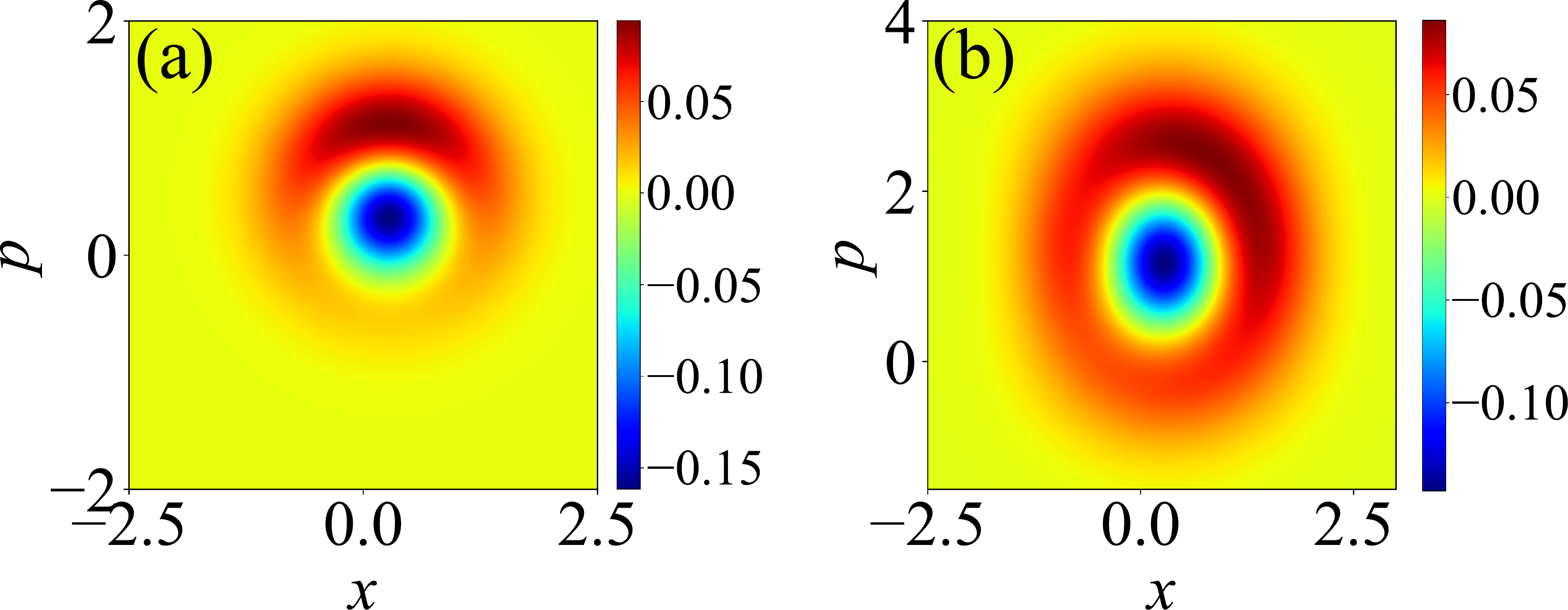}
    \caption{Calculated Wigner functions $W(x, p)$ of the fundamental driving mode when conditioned on ATI processes. (a) Conditioned on all electron momenta (Eq.~\eqref{eq:cat_ATI_cont}). (b) Conditioned on specific electron momentum (Eq.~\eqref{Eq:Sol:ATI}). Reproduced from Ref.~\cite{RLP22}.}
    \label{Fig:QED:ATI:Wigners}
\end{figure}

Figure~\ref{Fig:QED:ATI:Wigners}, shows representative examples of the Wigner functions of the state of the fundamental driving mode in case of conditioning on all (Fig.~\ref{Fig:QED:ATI:Wigners}a) and specific electron momenta (Fig.~\ref{Fig:QED:ATI}b), respectively. In the case of conditioning on all specific electron momenta, the shape of $W(x,p)$ in Fig.~\ref{Fig:QED:ATI:Wigners}a is very similar to an optical ``cat'' state generated when the IR field state is conditioned on HHG. This is valid under the following approximations: (i) during the ATI process the harmonic coherent-state amplitudes stay very close to the vacuum, and (ii) that the generated coherent shifts are identical and time-independent. However, the situation changes drastically in case of conditioning on specific electron momentum. In this case the $W(x,p)$ significantly deviates from the Wigner function of an optical ``cat'' state generated by conditioning on HHG. This is because the final state is given as a superposition of different coherent states (which, in principle, is larger than two), where each of them is affected by the instantaneous value of the electric field operator evaluated at time $t'$. The shape of $W(x,p)$ in Fig.~\ref{Fig:QED:ATI:Wigners}b is related to a change in the phase of the coherent states appearing in the superposition. However, it may also be associated with small rotations relevant to a change in the phase of the respective amplitudes in the state superposition, which at the end is related on how we are implementing the conditioning operations. Details on this matter can be found in Ref.~\cite{RLP22}.

\subsection{Experimental approach: Generation of optical ``cat'' states}
\label{ExpApproach}

\begin{figure*}
    \centering
    \includegraphics[width=1.0 \textwidth]{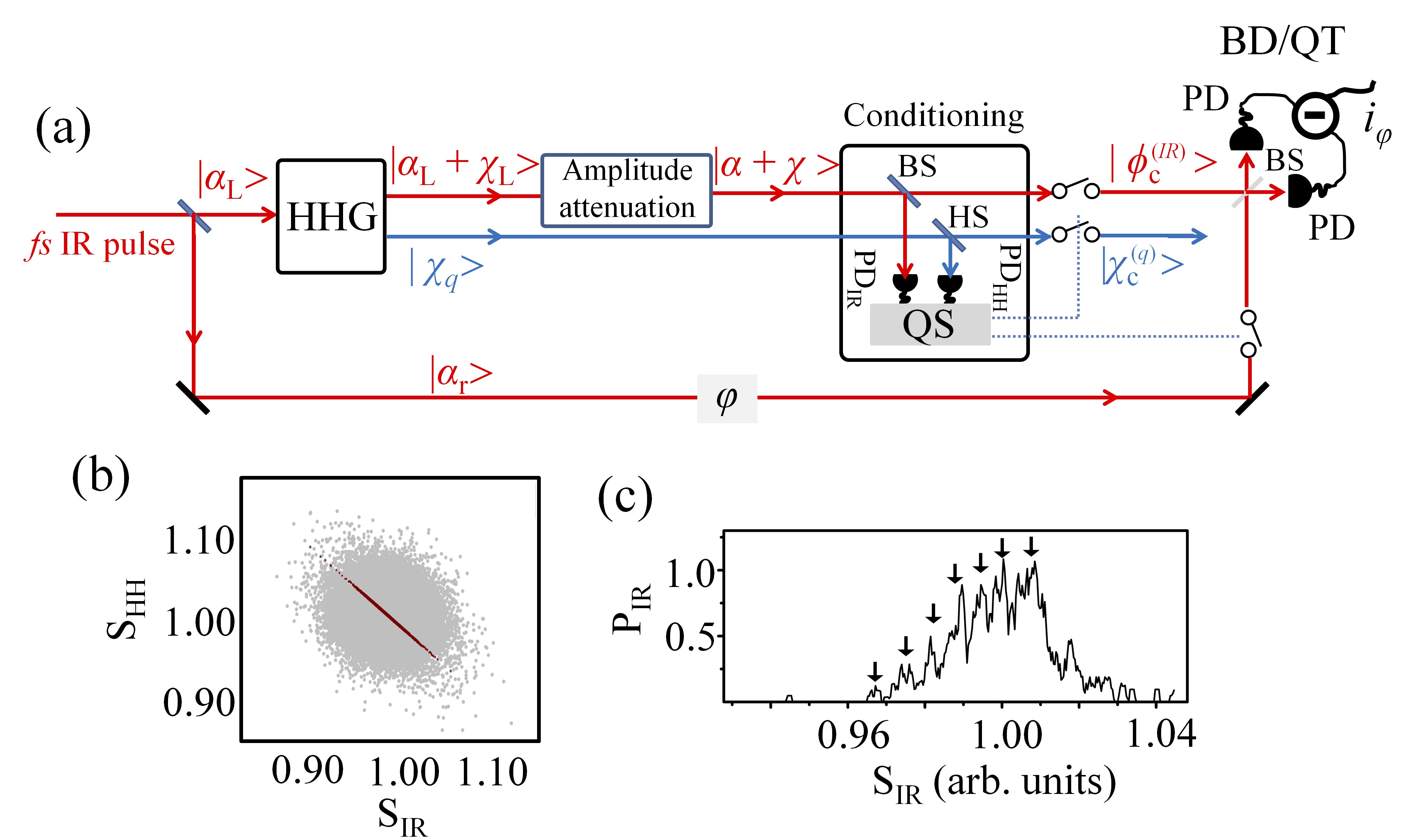}
    \caption{A schematic which shows the operation principle of the experimental approach. $\ket{\alpha_{L}}$ is the initial coherent state of the driving field. $\ket{\alpha_{L}+\chi_{L}}$, $\ket{\alpha + \chi}$ are the states of the IR field after the interaction and attenuation, respectively. $\ket{\chi_{q}}$ are the coherent states of the harmonic ($q$ is the harmonic order). $\ket{\phi_{c}^{(IR)}}$ and $\ket{\chi_{c}^{(q)}}$ are the field states conditioned on  HHG, which correspond to a coherent state superposition. BD/QT is a balanced detector (BD) of the homodyne detection system used for the characterization of the state $\ket{\phi_c}$ via quantum tomography (QT). $i_{\phi}$ is the $\phi$ dependent output photocurrent difference used for the measurement of the electric field operator and the light state characterization via the reconstruction of the Wigner function. $\ket{\alpha_r}$ is the state of the local oscillator reference field, with a controllable phase shift $\phi$. PD are identical IR photodiodes and PD$_{HH}$ is an HH photodetector. BS is a beam splitter. $S_\textrm{IR}$ and $S_\textrm{HH}$ are the signals of the PD and PD$_{HH}$, used for the conditioning on HHG.  HS is a plane mirror, named harmonic separator. HS, in combination with spectral filtering optical elements, can be used to reflect either all or part of the high harmonics towards PD$_{HH}$. QS is the quantum spectrometer approach used to condition the field states on the HHG process. (b) $(S_\textrm{HH}, S_\textrm{IR})$ photon number distribution (gray points). The mean values of $(S_\textrm{HH}$ and $S_\textrm{IR})$ are normalized to 1. The red points show the selected points along the anticorrelation diagonal. (c) Probability of absorbing IR photons ($\textrm{P}_{IR}$) towards HHG. The arrows depict the positions of the peaks in the multi--peak structure of $\textrm{P}_{IR}$. (b) and (c) reproduced from Ref.~\cite{SRM2023}. }
    \label{Fig:ExpSetup}
\end{figure*}

In this subsection we will discuss the operation principle of an experimental approach that can be used for the implementation of the aforementioned theoretical findings. The scheme, shown in Fig.~\ref{Fig:ExpSetup}a, allows for: (a) the generation of the non--classical light states by implementing conditioning approaches on the field modes after the interaction, (b) the control of the quantum features of the generated non--classical light states, and (c) the characterization of the quantum states of light. The approach has been successfully used for the generation and the characterization of high photon number controllable optical ``cat'' states using strongly laser driven atoms and conditioning on HHG \cite{LCP21,RLP22,SRM2023}.

\begin{figure*}
    \centering
    \includegraphics[width=0.4 \textwidth]{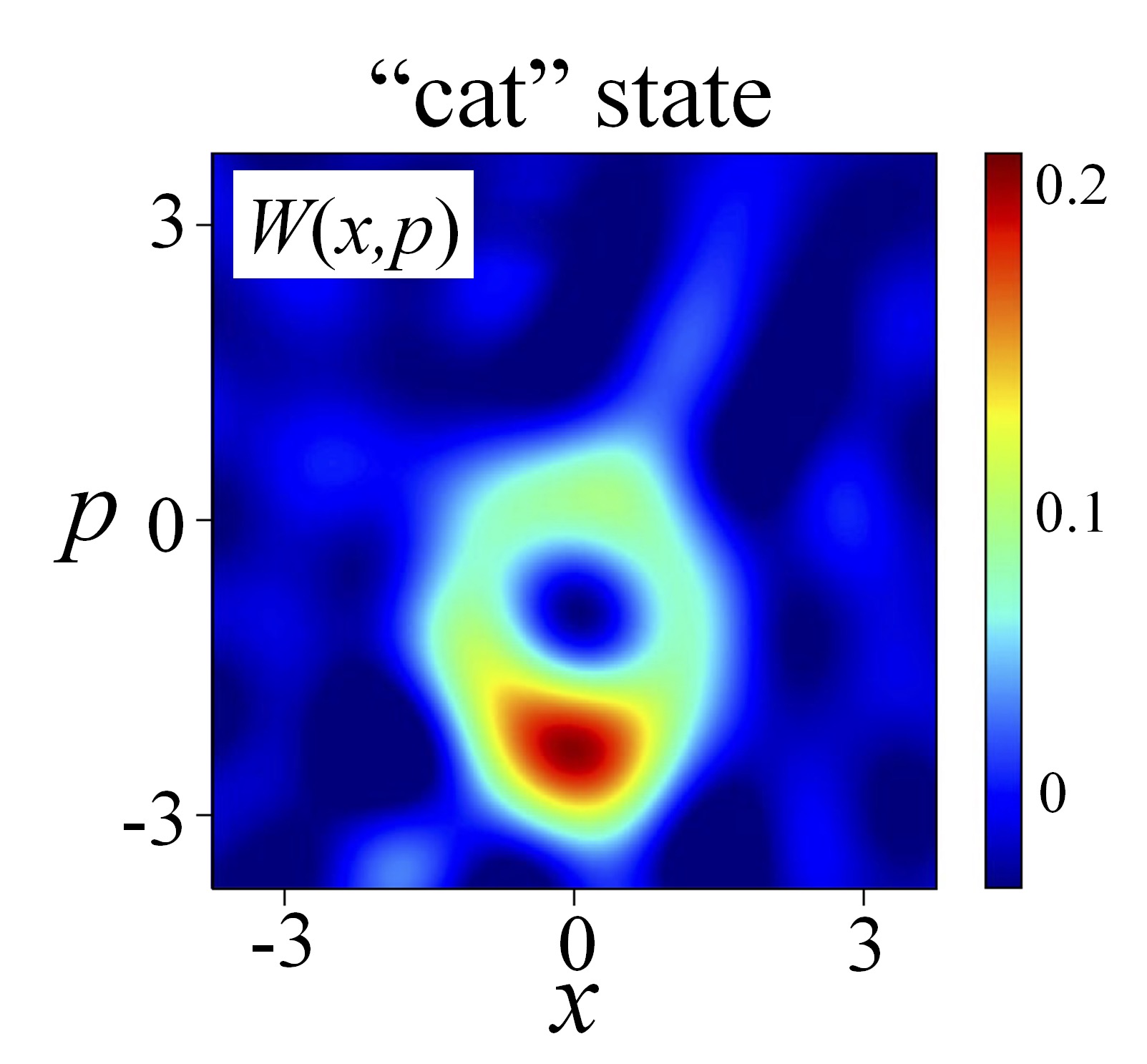}
    \caption{Measured Wigner function $W(x,p)$ of an IR optical ``cat'' state generated by conditioning on the HHG process. The measurement is in agreement with the Wigner functions obtained by theoretical calculations of a ``cat'' state with $|\alpha|=2$ and $|\chi|=1.5$ where $\xi \approx 0.32$.  Reproduced from Ref.~\cite{LCP21}.}
    \label{Fig:ExpCAT}
\end{figure*}

Briefly, a linearly polarized IR fs laser pulse is focused with an intensity $I_{IR} \sim 10^{14}$ W/cm$^2$ into an atomic gas leading to the generation of high harmonic photons, ions and photoelectrons (not shown in the Fig.~\ref{Fig:ExpSetup}a). The coherent state of the driving laser field before the interaction is $\ket{\alpha_L}$. The state of the field during the interaction with uncorrelated atoms (as are the atoms in a room temperature gas medium) remains coherent, but is modified with an amplitude shift to lower values. Therefore, the state of driving field after the interaction is an amplitude-shifted state $\ket{\alpha_{L}+\chi_{L}}$ and the generated harmonics are in coherent state $\ket{\chi_{q}}$. 
Then, the mode of the driving field is attenuated (resulting to $\ket{\alpha_{L}+\chi}$) in the range of few--photons and conditioned
to HHG by means of quantum spectrometer (QS) method \cite{GTK2016, TKG17, TKD19, LCP21, RLP22, SRM2023}. The
QS is a shot--to--shot IR vs HH photon correlation--based method, which selects
only the IR shots that are relevant to the harmonic emission (Fig.~\ref{Fig:ExpSetup}b). It relies
on photon statistics measurements and energy conservation, i.e., $q$ IR
photons need to be absorbed from the IR field for the generation of a photon
of the $qth$ harmonic. The output of the QS contains only the IR shots
along the anti--correlation diagonal (Fig.~\ref{Fig:ExpSetup}c). 
By selecting these points, we condition the IR field state exiting the medium on the HHG process. This is because  we select only those shots that are relevant to the harmonic emission, and we remove the unwanted background associated with all residual processes, e.g. electronic excitation or ionization. This action corresponds to the application of the $\Pi_{\Tilde{n}\neq 0} = \mathbbm{1}-\Pi_{\Tilde{0}}$ operator onto the field state given by Eq. \eqref{Eq:HHG:cond}, and its projection onto the harmonic field modes Eq.\eqref{Eq:HHG:Cond:IR}) \cite{SRL22,Sta22}. This results in the creation of an IR field coherent state superposition  (Eq. \eqref{Eq:HHG:Cond:IR}) i.e. an optical Schr\"{o}dinger ``cat'' state of the form,
\begin{equation}\label{Eq:Exp_HHG:Cond:IR}
    \ket{\phi_{c}^{(\text{IR})}}
        = \ket{\alpha_L + \chi}
            -\xi \ket{\alpha_L}.
\end{equation}
The quantum character of this state can be obtained by the measurement of its Wigner function. Figure~\ref{Fig:ExpCAT} shows the measured Wigner function of an IR optical ``cat'' state that has been generated by conditioning on the HHG. In this experiment, the high harmonics have been generated by the interaction of Xenon atoms with an $\approx 30$ fs IR pulse of intensity $I_{L}\approx 8\times 10^{13}$ W/cm$^2$. For the conditioning, the harmonics with $q > 11$ (i.e. mainly the plateau harmonics) have been used.

\subsubsection{Controlling the quantum features of the optical ``cat'' states}
\label{control}
\hfill \break

\begin{figure}
    \centering
    \includegraphics[width=0.8 \textwidth]{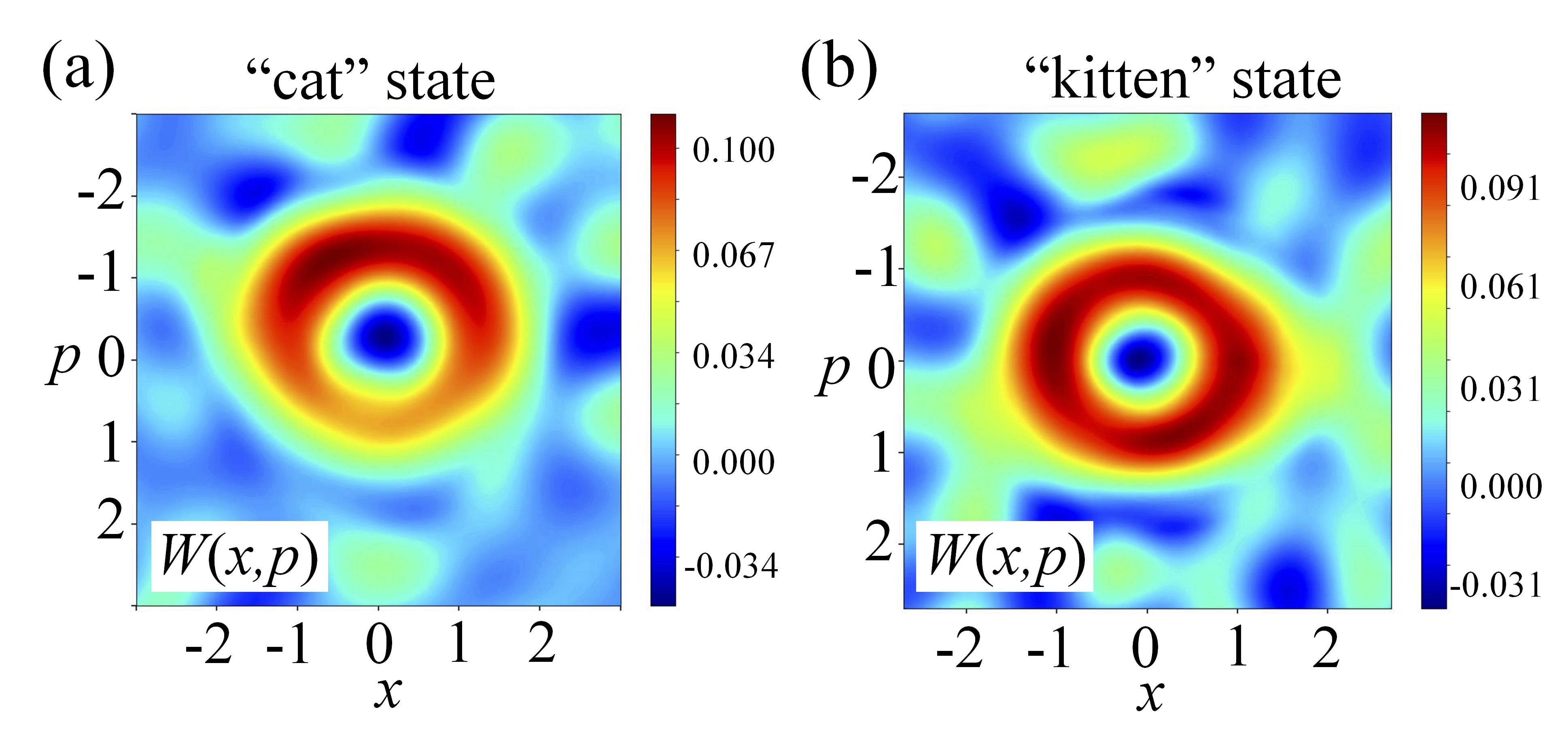}
    \caption{Measured Wigner functions $W(x, p)$ of an IR optical ``cat'' (a) and ``kitten'' (b) state, respectively. Optical ``cat'' and ``kitten'' states are created by conditioning on HHG for different values of $|\chi|$. The measurement is in agreement with the Wigner functions obtained by theoretical calculations of a ``cat'' state with $|\alpha|=1.4$ and $|\chi|=0.5$ where $\xi \approx 0.88$ and a ``kitten'' state with $|\alpha|=1.3$ and $|\chi|=0.1$ where $\xi \approx 0.99$. Reproduced from Ref.~\cite{RLP22}.}
    \label{Fig:Control:CAT}
\end{figure}

The experimental approach discussed in Sec.~\ref{ExpApproach}, has been used for the generation of high--photon number shifted
optical ``cat'' states and coherent state superposition with controllable quantum features using intense laser--atom interactions \cite{LCP21, RLP22, SRM2023}. The control of the quantum features is shown in Fig.~\ref{Fig:Control:CAT}a and Fig.~\ref{Fig:Control:CAT}a, where an optical ``cat'' state switches to a ``kitten'' for lower values of $\chi$. The value of $\chi$ can be controlled by changing the gas pressure in the interaction area (Eq.~\eqref{Eq:NrAtoms}). As is shown in Sec.~\ref{conditioningATI} the same approach can be used for conditioning on ATI \cite{RLP22, SRM2023}, and, in principle, can be extended to two-electron ionization processes, etc. This, besides its fundamental interest associated with electron-photon correlation during the ATI/HHG process, provides an additional ``knob'' of controlling the quantum character of the optical ``cat'' states, a feature extremely valuable for applications in quantum technology. Also, as it was discussed in Sec.~\ref{conditioningHHG_2colors}, single and two-color driven laser-atom interactions can result in the generation of ``massively'' entangled optical coherent states in the spectral range from extreme ultraviolet (XUV) to the far--infrared. Such states in combination with passive linear optical elements (such as phase shifters, beam splitters, and optical fibers) \cite{SRM2023} can be considered as unique resources for novel applications in quantum technology.

\subsection{Extension to interactions with complex materials and generation of massively entangled states}
\label{HHGmaterials}
The process of HHG has been observed when a strong laser field drives all states of matter. It has been observed in strongly laser driven molecules, clusters, liquids, semiconductor materials, nanostructures, 2-D materials, solid-surfaces via laser--plasma interactions \cite{Chatziathanasiou2017, ABC19, Ciappina2017, Nayak2019}, and very recently in quantum correlated materials such as high temperature superconductors \cite{ABB22}. The arrangement used for investigating these interactions is based on the schematic shown in Fig.~\ref{fig:Fig_Intense_laser_matter}. It is noted that in case of interactions with solid surfaces \cite{Chatziathanasiou2017, Nayak2019}, the driving laser field and the high harmonics are reflected from the target area (not shown in Fig.~\ref{fig:Fig_Intense_laser_matter}). The intensity of the laser pulse in the interaction region depends on the properties of the target. Typically is $I_{L}>10^{11}$ W/cm$^2$ for solid state and 2D materials, $I_{L}>10^{14}$ W/cm$^2$ for atoms (such as Noble gases) and molecules (mostly diatomic), and $I_{L}>10^{18}$ W/cm$^2$ for laser--plasma interactions on solid surfaces. 

In the majority of the aforementioned interactions, the HHG process has been described by semi-classical methods, which in many cases rely on the extension/modification of the methods developed for the description of intense laser--atom interaction (see Sec.~\ref{Sec:SLFP:SemiClas}). Although the fully quantized description of these interactions requires a detailed study (for example in atomic and molecular targets one can use TDSE or SFA \cite{ABC19}, for solids sometimes semiconductor Bloch equations, and for strongly correlated systems more sophisticated methods), it can be considered that, in principle, they have the potential to serve as additional resources for engineering non-classical and entangled states using the approach described in Sec.~\ref{ExpApproach}. This is because in all these interactions: 

\begin{itemize}
\item The driving field before the interaction is in a coherent state $\ket{\alpha_{L}}$. 

\item After the interaction, the energy conservation suggests that the state of the driving field mode is an amplitude-shifted coherent state $\ket{\alpha_{L}+\chi}$, and the emitted high harmonics are in a coherent state $\ket{\chi_{q}}$. We note that in the case of interactions with quantum correlated materials the outgoing for interaction field modes (fundamental and harmonic) could be in non--classical states \cite{PGR23}. 

\item After the interaction, the fundamental laser field can be attenuated in a coherent manner in the range of few--photons and conditioned to HHG by means of quantum spectrometer (QS). This will result in the creation of non-classical light states.
\end{itemize}

Such investigations have been already started in strongly laser driven semiconductors \cite{Gonoskov2022,RSM23}. In this situation, an electron may tunnel our from a parent atom (a parent site in the lattice) and recombines on another atom (another site). If the dynamics on the valence band is slow, this situation rarely happens and the physics is very similar to that of atoms. {\it Quantumness} can be generated using conditioning. If the valence intraband dynamics is faster, it may lead to noticeable entanglement between electron final position and light and non-classical states of light without the need for conditioning, which, however may augment the quantum effects. If the valence intraband dynamics is too fast, electron may recombine anywhere with random phase, which leads to strong decoherence and the impossibility of efficient phase matching of harmonics \cite{BJS22}.

Also theoretical investigations in strongly laser driven quantum correlated materials \cite{PGR23, SBA18}, and proposals towards the implementation of the approach in relativistic laser plasma interactions \cite{LLH21} have been reported. Furthermore, recent developments on the generation of high intensity squeezed light sources \cite{Manceau2019, Spasibko2017, Mouloudakis2020, Lamprou2020} have initiated experimental and theoretical investigations on the influence of the quantum features of these light states on the multi-photon processes \cite{Spasibko2017, Mouloudakis2020}, as well as, theoretical studies on the influence of the quantum noise of a squeezed light field on the HHG process \cite{TBG22}.

Taking into account the high photon number of the driving field and the large number non--classical field--modes (fundamental and harmonics) generated using the aforementioned methods, it can be considered that the strongly laser--driven materials is an ideal resource for engineering massively quantum correlated states which is of fundamental importance for quantum information science and quantum technologies.

\section{Applications in quantum information science}

Contemporary quantum technologies face major difficulties in fault tolerant quantum computing with error correction, and focus instead on various shades of quantum simulation (Noisy Intermediate Scale Quantum, NISQ) devices \cite{Preskill18}, analog and digital quantum simulators \cite{GAN14} (for a recent report on progress  including quantum computing and quantum simulation, see \cite{lnp1000}), and quantum annealers \cite{FFG01}. There is a clear need and quest for systems that, without necessarily simulating quantum dynamics of some physical systems, can generate massive, controllable, robust entangled and superposition states. This will, in particular, allow for the use of decoherence in a controlled manner, enabling the use of these states for quantum communications \cite{GT07} (e.g. to achieve efficient transfer of information in a safer and quicker way), quantum metrology \cite{Giovannetti2011}, sensing and diagnostics \cite{DRC17} (e.g. to precisely measure phase shifts of light fields, or to diagnose quantum materials).

As we discuss in this report, QED of strong laser physics, combined with QO, leads to non-trivial applications in QI science. Indeed, it offers a set of stable and reproducible methods to generate massively entangled states and massive quantum superpositions \cite{LCP21, RLP22, SRL22, SRM2023}. In this section we discuss that in fact there are more new paths for QI science via the symbiosis with AP and QO. These studies concern, in the first place, the fundamental QI science, but aim at quantum technologies. They deal in particular with: i) Detection of topology, strongly correlated systems, chirality, etc. (cf. \cite{ABB22,BBG22}); ii) Generation of topology, strongly correlated systems, chirality, etc. (cf. \cite{BCG22}); iii) Generation of entangled/quantum correlated states using conditioning methods (cf. \cite{LCP21,RLP22,SRM2023,RSM23});  iv) SLP and AP driven by non-classical light; v) Studies  of  quantum correlated states/decoherence in {\it Zerfall} processes (cf. \cite{maxwell_entanglement_2022}). 

\subsection{Detection of topology, strongly correlated systems, chirality, etc.}

Recently, there is an explosion of interest in applying AP methods to diagnose and detect topological order, strongly correlated systems, and chirality. The primary role is played here by the process of high harmonic generation (HHG). This is important for QI science at least for two reasons: i) combined with conditioning methods, this will definitely lead to novel ways of generating massive entanglement and massive quantum superpositions; ii) strongly correlated material exhibit themselves many-body entanglement; strongly correlated topological or chiral materials typically long range entanglement \cite{Wen-book} -- HHG will possibly allow to detect and characterize it. 

\subsubsection{Detection of topological order with HHG}
\hfill \break
\hfill \break
The pioneering work of R. Huber {\it et consortes} \cite{RSS18} studied lightwave-driven Dirac currents in a topological surface band. ``Harnessing the carrier wave of light as an alternating-current bias may enable electronics at optical clock rates'' \cite{RSS18,KS14,HLS15,GR19}. In these papers the authors observe directly how the carrier wave of THz light pulse accelerates Dirac fermions in the topological state of Bi$_2$Te$_3$. 

\begin{figure}[t]
    \centering
    \includegraphics[width = 1.0 \textwidth]{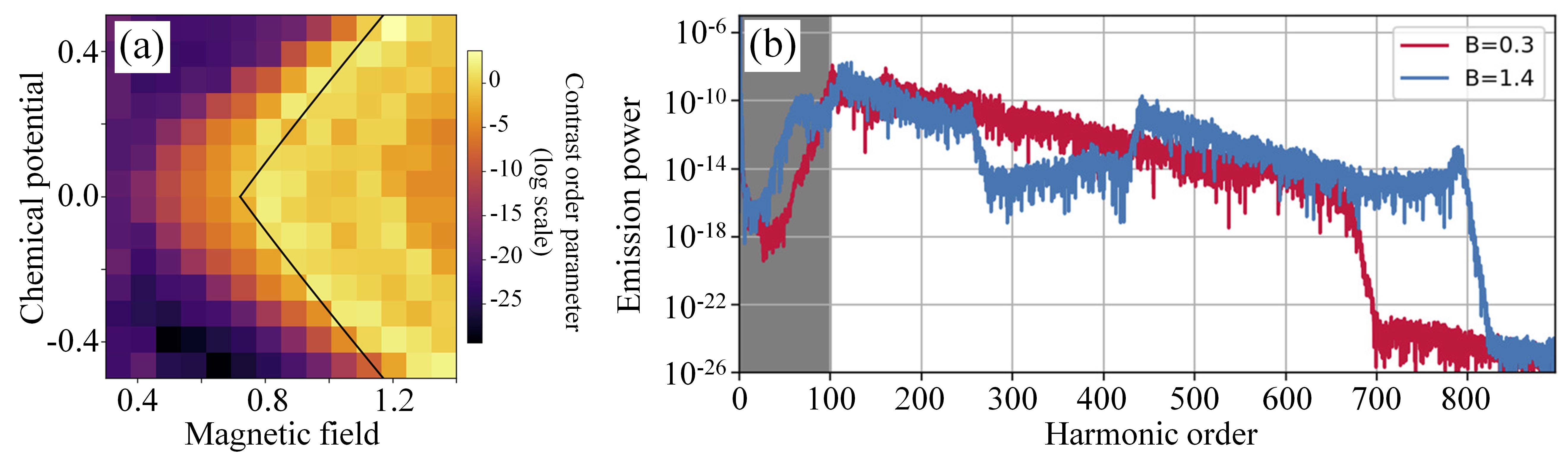}
    \caption{(a) The phase diagram of the proximity-induced p-wave superconductor is obtained from an order parameter based on the harmonic emission spectrum,  shown in (b) for two parameter choices, in the topological and in the trivial phase. At frequencies below the band gap, emission is suppressed in the trivial phase (red), whereas in the topological phase emission is possible already at frequencies above half the band gap due to the edge state in the middle of the band gap. Reproduced from ref. \cite{BBG22}.}
    \label{Fig:p-wave}
\end{figure}

The first papers, where HHG was used for the detection of topology, were the series of papers by D. Bauer with his collaborators on the detection of topological edge states in dimerized chains \cite{BH18,JB19}, first using a self-consistent time-dependent density functional theory approach. For harmonic photon energies smaller than the band gap, the harmonic yield was found to differ by 14 orders of magnitude for the two topological phases. Similar conclusions were obtained for  Su-Schrieffer-Heeger (SSH) chains  that display topological edge states. Recently, this method  has been extended to the detection of Majorana fermions in the Kitaev chain \cite{pattanayak21} and in quantum wires with proximity-induced p-wave superconductivity \cite{BBG22} (see Fig.~\ref{Fig:p-wave}). Specifically, the harmonic emission spectrum in strong fields was shown to exhibit spectral features due to radiating edge modes, present in the topological phase, and absent in the trivial phase. The results suggest HHG spectroscopy as a novel all-optical tool for the detection of, still controversial, Majorana zero modes.
 
In the seminal {\it Nature Photon.} paper, M. Ivanov {\it et consortes} discussed topological strong-field physics on sub-laser-cycle timescale \cite{SJA19}. These authors studied the response of the Haldane model \cite{Hal88} to ultrashort, ultrastrong laser pulses. They demonstrated that electrons tunnel differently in trivial and topological insulators due to the key role of the Berry curvature.  Attosecond delays and helicities of harmonics serve then to the phase diagram of the system. Strong fields can also be used for manipulations of topological properties of 2D materials, relevant for valleytronic devices \cite{JSS20}. A. Chacón {\it et al.} \cite{CZK20} obtained complementary results studying circular dichroism, which clearly heralded topological phases and transitions in  the Haldane model and other Chern insulators. 

\subsubsection{Detection of strongly correlated systems with HHG}
\hfill \break
\hfill \break
This is another rapidly developing area. As pointed out in  the recent review \cite{GB22}: ``A hallmark effect of extreme nonlinear optics is the emission of high-order harmonics of the laser from the bulk of materials. The discovery and detailed study of this phenomenon over the course of the past decade has offered a broad range of possibilities and seen the dawn of a new field of extreme solid-state photonics. In particular, it opens the way to previously inaccessible spectral ranges, as well as the development of novel solid-state spectroscopy and microscopy techniques that enable detailed probing of the electronic structure of solids.''

Pioneering studies of HHG in strongly correlated materials are described in the paper by Silva {\it et al.} \cite{SBA18}, where the HHG  is used to time-resolved ultrafast many-body dynamics of an optically driven phase transition to Mott insulator. Further studies on two-dimensional Mott insulators were carried out by Orthodoxou et al. using quantum Monte Carlo, exact diagonalization, and a simplified dynamical mean-field model~\cite{Orthodoxou2021}. The theory on HHG from Mott insulators was also reported recently by the Tokyo group, who used infinite time-evolving block decimation (iTEBD) method and exact diagonalization \cite{MTK21,MUK22}, and more recently, together with Ph.~Werner demonstrated anomalous temperature dependence of HHG in Mott insulators. This reveals the crucial effect of strong spin-charge coupling on HHG in Mott insulators. In a system with antiferromagnetic correlations, ``the HHG signal is drastically enhanced with decreasing temperature, even though the gap increases and the production of charge carriers is suppressed. This anomalous behavior, which has also been observed in recent HHG experiments on Ca$_2$RuO$_4$ \cite{UMY22},
originates from a  cooperative effect between the spin-charge coupling and the thermal ensemble, as well as the strongly temperature-dependent coherence between charge carriers'' A bosonic analog of fermionic HHG in a strong time-dependent synthetic vector potential has also been theoretically investigated by 
Roy et al. \cite{Roy2020} as a phase distinction measure to distinguish between the Mott and the superfluid phase. HHG has even been proposed to be used as a probe to identify dynamic features of hard to measure quantum spin liquids, as the magnetic-field dependence of the HHG spectra can drastically differ from those of usual ordered magnets \cite{Kanega2021}.

Ultrafast laser and HHG based solid state spectroscopic techniques have increasingly become popular.  Polarization-state-resolved high-harmonic spectroscopy of solids was reported in Ref.~\cite{Klemke2019}. Ultrafast pump-probe spectroscopy has been used to detect the presence of Hubbard excitons in certain Mott insulators \cite{lovinger2020influence}. Femtosecond optical pulses have also been used to stabilize non-thermal transient phases lasting few picoseconds in the Mott-Hubbard insulator di-Vanadium trioxide \cite{Lantz2017}. In Ref. \cite{BHL21}, the authors establish time-resolved high harmonic generation for tracking photo-induced dynamics of the insulator-to-metal phase transitions in vanadium dioxide.
Closely, related to this is the study by Johnson et al. where they use time and spectrally resolved coherent X-ray imaging to track the light-induced insulator-to-metal phase transition in vanadium dioxide on the nanoscale with femtosecond time resolution\cite{Johnson2022} and also observed the ultrafast loss of lattice coherence in the light induced structural phase transition of di-Vanadium trioxide \cite{Johnson2022prl}.

Studies of the detection of Majorana fermions \cite{BBG22,pattanayak21} should also be mentioned in this context.  MBI groups of Ivanov and Smirnova studied the strong laser field response of  the two-band Hubbard model, using advanced techniques of dynamical mean field theory (\cite{VGE22}, see also  M. Ivanov's contribution to \cite{ATTOVIII}).  In this work, the authors  introduced a new type of non-linear approach that allows  unraveling the sub-cycle dynamics of strongly correlated systems interacting with few-cycle infrared pulses. Their approach can resolve pathways of charge and energy flow between localized and delocalized many-body states in the 2D Hubbard model; it allows also to describe  the creation of highly correlated states that survive after the end of the laser pulse. 

Other applications of HHG to strongly correlated materials and condensed matter systems include: i) extraction of higher-order nonlinear electronic response in solids \cite{HOK19}; ii) HHG enhancement in solid state nano-structures, such as graphene heterostructures \cite{CRT22}; iii) characterization of quantum criticality in strongly correlated systems \cite{SLZ22}: In the latter work exact diagonalization method was applied for the extended Hubbard model on a periodic chain. It was shown that  HHG close to the quantum critical point, is more efficient than in the gapped charge-density-wave and spin-density wave phases; iv) Ohio group led by A. Landsmann \cite{AML22} studied ultrafast laser-driven dynamics in metal-insulator interface using HHG, and showed the field induced
dielectric break-down at the Mott-insulator/metal interface. They demonstrated that the intensity of high harmonic emission correlates closely with double production and the corresponding loss of short-range anti-ferromagnetic order.

 HHG has also studied in cuprates (YBCO) in a broad temperature range from 80K to 300K, probing different phases of the high-$T_c$ material  \cite{ABB22}. Amazingly, HHG differs in various strongly correlated phases: The superconducting phase is marked by a strong increase of emission at all odd harmonic frequencies (3rd, 5th, and 7th), whereas the transition from the pseudogap phase to the strange metallic phase is reflected by an intensity drop only in the highest harmonics (7th). The experimental results were  reproduced by simulating  a two-band model in the $d$-wave paring mean-field regime, including phenomenological scattering terms. In another study \cite{Vaswani2021}, using terahertz (THz) two-pulse coherent spectroscopy, coherent oscillation of the complex amplitude mode or Higgs mode has been observed in an iron-based superconductor, thus paving the way for light-based control of the Higgs mode, an illusive mode which is notoriously difficult to detect due to its non-linearly coupling with the electromagnetic field.

\begin{figure*}
    \centering
    \includegraphics[width=1.0 \linewidth]{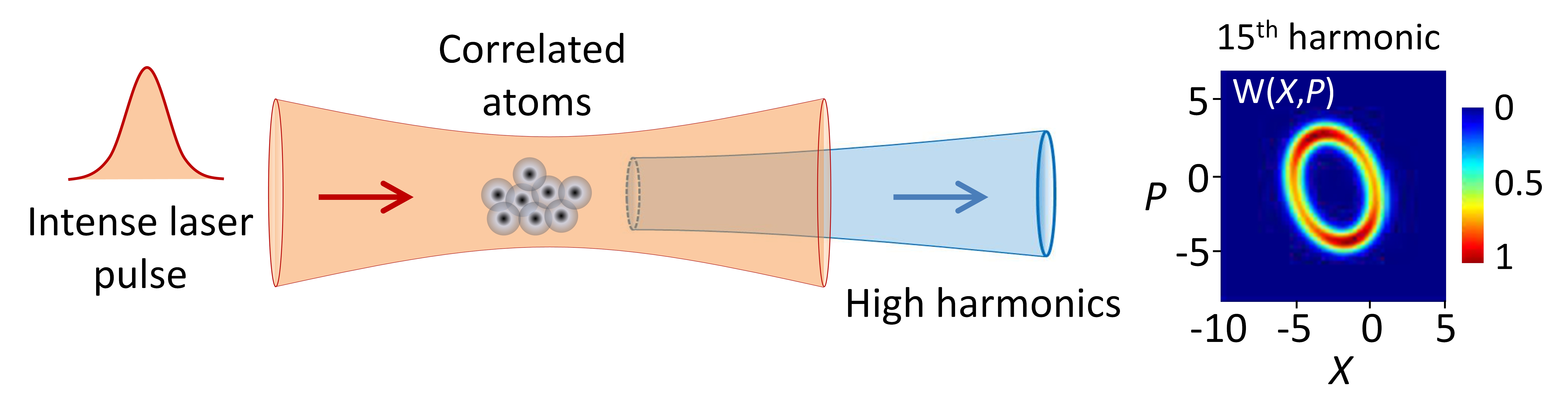}
    \caption{Strongly laser-driven quantum correlated many-body systems lead to the generation of light with exotic quantum features - the quantumness of a many-body system is imprinted on the state of the emitted light. Figure reproduced from ref. \cite{Tzallas2023} which refers to the findings of ref. \cite{PGR23}.}
    \label{figure-paris}
\end{figure*}

Clearly, it is already possible to image and manipulate strongly correlated materials with HHG at optical rates for ultrashort laser pulses, far beyond  Floquet engineering of quantum systems. For future QI applications, it is important to remember that in all these processes it is natural to expect the generation of massively entangled states of matter, light, and light-matter. Indeed, recently Pizzi {\it et al.} \cite{PGR23} proposed a quantum-optical theory of a strongly driven many-body system, showing that the presence of correlations among the emitters creates emission of many-photon states of light which deviate from the coherent states of light. The authors considered a simplified model of matter, described by a quadratic Hamiltonian, and took as  an example HHG. They demonstrated that a correlation of the emitters prior to the strong drive is converted onto features different from coherent radiation of the output light, including doubly-peaked photon statistics, ring-shaped Wigner functions, and correlations between harmonics (for a description for a general audience, see \cite{Tzallas2023} and Fig. (\ref{figure-paris})). However, genuine non-classicality in the properties of the emitted radiation is still not unambiguously present.

\subsubsection{Detection of chirality with HHG}   
\hfill \break
\hfill \break
The detection of chirality is somewhat similar to the detection of topological order with HHG, but nowadays mostly refers to molecular targets. Chirality is generic for biological molecules;  distinguishing molecules which differ only by their chirality has a gigantic technological importance which cannot be overstated. 
The relevance for QI science concern here additionally two aspects: i) Molecules (chiral or not) offer novel ways of generating entangled states, similar to those discussed for solids  
\cite{RSM23}, where an electron localized close to one nucleus, recombines on the other; ii) possibility of generalizing chirality imaging to many-body chiral systems.

In recent years, O.\ Smirnova and collaborators obtained numerous seminal results concerning ultrafast imaging of chiral molecules
(cf. \cite{CBP15}, see \cite{AOS22} for a recent review). In Refs. \cite{AN19, AO21} the authors  introduced the concept of \emph{locally chiral light}, i.e. light that is chiral within the electric-dipole approximation by means of its 3D chiral Lissajous figure. This type of chiral light can be synthesized and exploited to dramatically enhance the contrast between HHG spectra in molecules with opposite chiralities.  In Ref. \cite{AOI21}, they study the transverse spin of the light, which emerges whenever light is confined to a small region of space in ultra-fast chiral imaging. This results in harmonics with a polarization that depends on the molecular chirality. The effects of synthetic chiral light also extend to other attosecond-science properties, such as the use of a chirally-sensitive nonlinear Stark shift to deflect free-induction decay radiation in an enantiosensitive fashion \cite{Khokhlova2022}.

Recently, we found \cite{POL22} that chiral molecules subject to strong-field ionization with a few-cycle, IR, linearly polarized pulse emit \emph{twisted} photoelectrons with the twist (orbital angular momentum (OAM)) depending on the molecular chirality. This effect, termed {\it photoelectron vortex dichroism (PEVD)}, offers novel ways of imaging chirality, and suggests intriguing perspectives regarding the role of the electron OAM in recollision-based phenomena in chiral molecules.

\subsection{Generation of topology, strongly correlated systems, chirality, etc.}

As discussed above, AP and HHG serve as great detection tools; amazingly,  ultrafast laser pulse and attophysics methods may also be used for the generation of topological order, strongly correlated states, chirality, etc. In all these cases we foresee the possibility of the generation of novel entangled states of light, matter, and light-matter, relevant for QI science. The ideas reported here are, of course, close to the attempts to generate room temperature high $T_c$ superconductivity \cite{CavalieriHighT_c}, or Chern insulators \cite{MSS20}, which belong more to the domain of ultrafast, but THz physics. 

A good example of laser-induced phase transitions is discussed in the  experimental group of S. Wall on THz field control of in-plane orbital order in La$_{0.5}$Sr$_{1.5}$MnO$_4$ \cite{MCT15}. We have also started to investigate possibilities of laser induced fluctuating bonds in superconductivity \cite{JDC20} in hole-doped cuprates. In the recent Phys. Rev. B Lett. \cite{BCG22}, demonstrating the theoretical  possibility of generation of fermionic Chern insulators from twisted light with linear polarization in graphene-like materials. 

A particularly interesting line of research, connects generation of topology, strongly correlated systems, chirality etc., with the use of more complex {\it structured laser light}, combining polarization and OAM effects. A perfect example is a superposition of  left and right circular polarized two-colour laser fields in a, so-called, MAZEL-TOV configuration \cite{FKD14,KBH16}. Such driving leads to the famous three-foil Lissajous patterns of the electric field, and efficient generation of circularly polarized high harmonics.  

Another example is the light, which forms fractional-order knots in the dynamics of the electric field vector employing  the polarization state of light and superposition of the fundamental and doubled frequency \cite{PJV19}. Application of strong laser pulses of this form to atomic targets leads to ``exotic'' conservation of torus-knot angular momentum in high-order harmonic generation \cite{PRS19}.  Combining two delayed circularly polarized pulses of frequency $\omega$ and $2\omega$, one can generate light with a self-torque: extreme-ultraviolet beams of HHG with time-varying orbital angular momentum \cite{RDB19}. Combining structured laser light with conditioning opens new possibilities of generating massive entanglement and superpositions with a topological ``touch'' -- so far not yet explored areas for QI application.

As shown in Ref. \cite{MPM22}, \emph{locally chiral light} \cite{AN19, AO21}, another type of structured light, can be used to efficiently (i.e. at the level of the electric-dipole approximation) imprint 3D chirality on achiral matter, such as atoms. Locally chiral light sculpts a chiral orbital out of the initially achiral ground state, exciting the electron into an orbital with a chiral shape.

\subsection{Strong-field physics
and atto-second science driven by non-classical light}

All the quantum optical studies of strong field driven processes have thus far been approached by assuming a coherent state description for the driving field. The corresponding field is classical, even if Hilbert space methods are used for its description. 
However, in a recent work, presented at ATTO VIII by Even Tzur et al. \cite{Kaminer_squeezing}, this assumption was abandoned, and the process of HHG driven by non-classical light fields has been presented for the first time. In the recent paper \cite{GTB22} they show that the plateau and cutoff in HHG spectrum  are sensitive to the photon statistics of the driving light. While coherent (classical) and Fock light states induce the established HHG cutoff law, thermal and squeezed states substantially surpass it. This opens the path for investigating the interplay between non-classical properties of the driving source, and quantum properties of the harmonic modes. In another paper \cite{TBG22} they also  show, using a quantum generalization of SFA,  that dynamics of matter driven by bright (intense) light  significantly depend on the quantum state of the driving light, which induces an effective photon-statistics force. It is worth mentioning that these works go back to an earlier paper on ``Compton Scattering Driven by Quantum Light'' \cite{KK22}, where the authors obtain analytical results for the Compton emission spectrum when driven by thermal and squeezed vacuum states, showing a noticeable broadening of the emission spectrum relative to a coherent drive, thus reaching higher emission frequencies for the same average intensity.

\subsection{Generation of entanglement in {\it Zerfall} processes}

\begin{figure}
    \centering
    \includegraphics[width = 0.9 \textwidth]{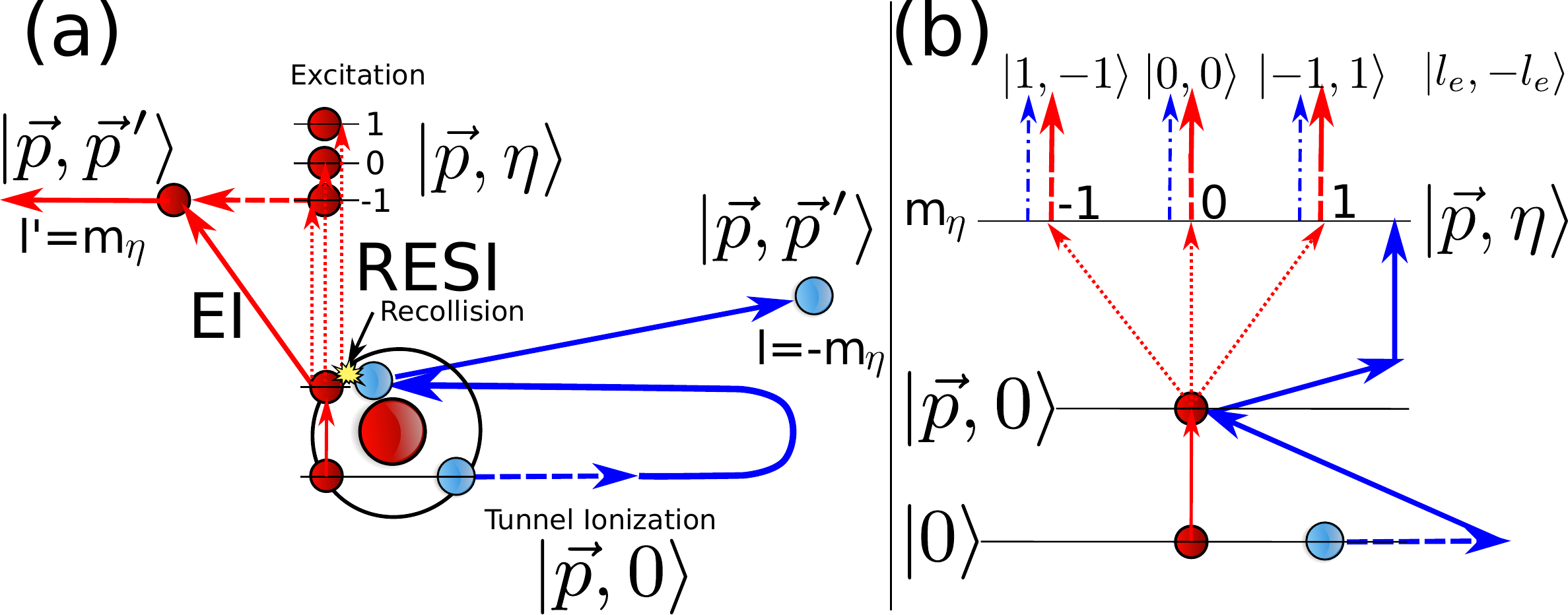}
    \caption{Non-sequential double ionization (NSDI), a `Zerfall' processes, leading to entanglement between the orbital angular momentum (OAM) of the two photoelectrons. In panel (a), an electron is ionized via a strong laser field and recollides with the parent atom/molecule. This leads to either: direct ionization of a second electron in electron impact (EI) ionization, or excitation with subsequent ionization (RESI) of a second electron. Panel (b) focuses on RESI, the OAM superposition in the excited state is transferred to the final state along with anti-correlation between the electrons. The figure is reproduced from ref.~\cite{lewenstein_attosecond_2022} reporting on result from ref.~\cite{maxwell_entanglement_2022}, which contained a similar figure.}
    \label{Fig:NSDI}
\end{figure}

Finally, AP is a perfect playground to study the generation of quantum correlations and entanglement in {\it Zerfall} processes, i.e. processes of decay of a “whole” in products. A prime example of this is non-sequential double ionization (NSDI), where strong-field ionization followed by laser induced recollision leads to double ionization of the target, see Fig.~\ref{Fig:NSDI}a. Strong correlation between the ionization products, in particular the two photoelectrons, has long been known \cite{weber_correlated_2000}. Previously, coherence and interference between the two photoelectrons has been studied in great detail \cite{MF15,MF16,HCL14, QHW17}. More recently, A. Maxwell {\it et al.} \cite{maxwell_entanglement_2022} studied generation of entanglement in non-sequential double ionization, by using correlation in the OAM of the outgoing electrons, for a mini-review on OAM in strong-field ionization see \cite{MAC21,KPC21}. A key step, is to exploit the intermediate excited state, and its inherent superposition over OAM, which is present for the `second' electron for the low-intensity regime\footnote{This is known as the recollision with subsequent ionization (RESI) mechanism.}. Thus, simple conservation laws dictate that the final OAM of the two electrons must be anti-correlated, which at certain final momentum leads to a maximally entangled qutrit, see Fig.~\ref{Fig:NSDI}b, while remaining robust to incoherent effects, such as focal averaging or decoherence with the ion. 

Continuum products in a \textit{typical} strong-field Zerfall processes will be quantified through continuous variables, which can complicate analysis of entanglement measures. However, the use of OAM makes the task of quantifying and measuring entanglement much easier, the computations of the reduced density matrix of ionized product in the OAM space become simple, and in the case of NSDI, clearly exhibit entanglement. This also enables quantification by the, so-called, logarithmic negativity, which may be used for mixed states to model incoherent effects. \\

\subsection{Characterizing decoherence in {\it Zerfall} processes}

Finally, studies of {\it Zerfall} processes offer a unique opportunity to understand and characterize quantum and classical decoherence. This goes back to seminal and pioneering theoretical works by 
R. Santra {\it et al.} \cite{PGH11,AVS17,ALK20},  F. Martín {\it et al.} \cite{LAT16}, or M. Vacher {\it et al.} \cite{VBR17}. All of these works dealt essentially with the single electron ionization process that might lead to entanglement of the electron with the parent ion. Measuring the reduced density matrix of the electron provides thus information about decoherence and thus entanglement
Mark Vrakking {\it et al.} has developed both theoretical and the first experimental ideas in this context \cite{Vra21,KMD22} (see also M. Vrakking's contribution to \cite{ATTOVIII}). 
One of the main problems in this line of research is to make sure that the source of decoherence is truly quantum. Indeed, by analyzing with great care various classical and quantum models of decoherence in the process of single electron ionization, Ch. Bourassin-Bouchet {\it et al.} demonstrated dominantly classical sources of decoherence. 

In contrast, in the recent work \cite{BLF22}, A. L'Huillier {\it et al.} investigated
decoherence due to entanglement of the radial and angular degrees of freedom of the photoelectron.
They study two-photon ionization via the $2s2p$ autoionizing state in He using high spectral resolution
photoelectron interferometry. Combining experiment and theory, we show that the strong dipole
coupling of the $2s2p$ and $2p^2$ states results in the entanglement of the angular and radial degrees
of freedom. This translates, in angle-integrated measurements, into a dynamic loss of coherence
during autoionization.

\section{Conclusions}
We report on the recent progress on the fully quantized description of intense laser--matter interactions and the development of approaches that have been used for the generation of controllable entangled and non-classical light states. This has led to the connection of strong laser field physics with quantum optics and quantum information science. Specifically, after an introduction to the fundamentals of quantum optics, non--classical light engineering and intense laser--matter interactions, we discuss the fully quantized description of intense laser--matter interaction emphasizing on quantum electrodynamics of strongly laser--driven atoms, high harmonic generation and above threshold ionization processes. Then, we discuss the quantum operations (conditioning) used to engineer optical ``cat'' and entangled states from the XUV to the far--IR, as well as the procedures that can be followed to control their quantum features. Additionally, we provide the perspectives towards the extension to interactions with complex materials as well as interactions with intense quantum light, that can serve as additional resources for engineering novel non-classical and entangled states. Finally, we discuss a number of applications stemming from the symbiosis of strong laser–field physics and ultrafast science with quantum optics and quantum information science.

\section*{Acknowledgements}
P. Tzallas group at FORTH acknowledges support from: LASERLABEUROPE V (H2020-EU.1.4.1.2 grant no.871124), FORTH Synergy Grant AgiIDA (grand no. 00133), the H2020 framework program for research and innovation under the NEP-Europe-Pilot project (no. 101007417) and ELI--ALPS. ELI--ALPS is supported by the European Union and co-financed by the European Regional Development Fund (GINOP Grant No. 2.3.6-15-2015-00001).

M. Lewenstein group at ICFO acknowledges support from: ERC AdG NOQIA; Ministerio de Ciencia y Innovation Agencia Estatal de Investigaciones (PGC2018-097027-B-I00/10.13039/501100011033,  CEX2019-000910-S/10.13039/501100011033, Plan National FIDEUA PID2019-106901GB-I00, FPI, QUANTERA MAQS PCI2019-111828-2, QUANTERA DYNAMITE PCI2022-132919,  Proyectos de I+D+I “Retos Colaboración” QUSPIN RTC2019-007196-7); MICIIN with funding from European Union NextGenerationEU(PRTR-C17.I1) and by Generalitat de Catalunya;  Fundació Cellex; Fundació Mir-Puig; Generalitat de Catalunya (European Social Fund FEDER and CERCA program, AGAUR Grant No. 2021 SGR 01452, QuantumCAT \ U16-011424, co-funded by ERDF Operational Program of Catalonia 2014-2020); Barcelona Supercomputing Center MareNostrum (FI-2022-1-0042); EU Horizon 2020 FET-OPEN OPTOlogic (Grant No 899794); EU Horizon Europe Program (Grant Agreement 101080086 — NeQST), National Science Centre, Poland (Symfonia Grant No. 2016/20/W/ST4/00314); ICFO Internal “QuantumGaudi” project; European Union’s Horizon 2020 research and innovation program under the Marie-Skłodowska-Curie grant agreement No 101029393 (STREDCH) and No 847648  (“La Caixa” Junior Leaders fellowships ID100010434: LCF/BQ/PI19/11690013, LCF/BQ/PI20/11760031,  LCF/BQ/PR20/11770012, LCF/BQ/PR21/11840013). Views and opinions expressed in this work are, however, those of the author(s) only and do not necessarily reflect those of the European Union, European Climate, Infrastructure and Environment Executive Agency (CINEA), nor any other granting authority.  Neither the European Union nor any granting authority can be held responsible for them. 

M. F. Ciappina acknowledges financial support from: The Guangdong Province Science and Technology Major Project (Future functional materials under extreme conditions - 2021B0301030005).

P. Stammer acknowledges funding from: The European Union’s Horizon 2020
research and innovation programme under the Marie Skłodowska-Curie grant agreement No 847517. 

J. Rivera-Dean acknowledges support from: The Secretaria d'Universitats i Recerca del Departament d'Empresa i Coneixement de la Generalitat de Catalunya, as well as the European Social Fund (L'FSE inverteix en el teu futur)--FEDER. 

A. S. Maxwell acknowledges funding support from: The European Union’s Horizon 2020 research and innovation programme under the Marie Sk\l odowska-Curie grant agreement, SSFI No.\ 887153.

E. Pisanty acknowledges Royal Society University Research Fellowship funding under grant URF\textbackslash{}R1\textbackslash{}211390.

Authors contributions: U.~B, Th.~L., A.~M., A.~O., E.~P., J.~R.-D., P.~S.: Equally contributed authors placed in the author list in alphabetical order. M.~F.~C., M.~L.\ and P.~T.: Supervised the work.

\newcommand{\newblock}{}
\bibliographystyle{arthur}

\bibliography{ROPP.bib}

\end{document}